\documentclass[usenatbib,fleqn]{mnras}

\usepackage{aas_macros}
\usepackage{amsmath}
\usepackage{amssymb}
\usepackage{booktabs}
\usepackage{bm}
\usepackage{color}
\usepackage{textcomp}
\usepackage{gensymb}
\usepackage{graphicx}
\usepackage{hyperref}
\usepackage{lineno}
\usepackage{siunitx}
\usepackage{stackengine}
\usepackage{verbatim}

\newcommand{\xfigure}[1]{
\begin{center}
\includegraphics[width=0.5\textwidth]{figs/#1}
\end{center}
}

\newcommand{\be}{\begin{equation}}
\newcommand{\ee}{\end{equation}}

\newcommand{\bi}[1]{\textit{#1}}

\newcommand{\xinclude}[1]{\includegraphics[width=0.47\textwidth]{figs/#1}}

\title[The PAU Survey: Photometric redshifts]{The PAU Survey: Early
demonstration of photometric redshift performance in the COSMOS field}
\author[The PAU Survey: Photometric redshifts]{M. Eriksen$^{1}$\thanks{E-mail: eriksen@pic.es}\thanks{Also at Port d'Informaci\'{o} Cient\'{i}fica (PIC), Campus UAB, C. Albareda s/n, 08193 Bellaterra (Cerdanyola del Vall\`{e}s), Spain}, 
A. Alarcon$^{2,3}$,
E. Gaztanaga$^{2,3}$,
A. Amara$^{4}$,
L. Cabayol$^{1}$,
\newauthor
J. Carretero$^{1}\footnotemark[2]$,
F. J. Castander$^{2,3}$,
M. Crocce$^{2,3}$,
M. Delfino$^{1}\footnotemark[2]$,
J. De Vicente$^{5}$,
\newauthor
E. Fernandez$^{1}$,
P. Fosalba$^{2,3}$,
J. Garcia-Bellido$^{6}$,
H. Hildebrandt$^{7}$,
H. Hoekstra$^{8}$,
\newauthor
B. Joachimi$^{9}$,
P. Norberg$^{10}$,
R. Miquel$^{1,11}$, 
C. Padilla$^{1}$,
A. Refregier$^{4}$,
E. Sanchez$^{5}$, 
\newauthor
S. Serrano$^{2,3}$,
I. Sevilla-Noarbe$^{5}$,
P. Tallada$^{5}\footnotemark[2]$,
N. Tonello$^{1}\footnotemark[2]$,
L. Tortorelli$^{4}$ \\ \\
$^{1}$ Institut de F\'{\i}sica d'Altes Energies (IFAE), The Barcelona Institute of Science and Technology, 08193 Bellaterra (Barcelona), Spain \\
$^{2}$ Institute of Space Sciences (ICE, CSIC), Campus UAB, Carrer de Can Magrans, s/n, 08193 Barcelona, Spain \\
$^{3}$ Institut d'Estudis Espacials de Catalunya (IEEC), E-08034 Barcelona, Spain \\
$^{4}$ Institute for Particle Physics and Astrophysics, ETH Z\"urich, Wolfgang-Pauli-Str. 27, 8093 Zürich, Switzerland \\
$^{5}$ Centro de Investigaciones Energ\'eticas, Medioambientales y Tecnol\'ogicas (CIEMAT), Avenida Complutense 40, \\
28040 Madrid (Madrid), Spain \\
$^{6}$ Instituto de Fisica Teorica (IFT-UAM/CSIC), Universidad Autonoma de Madrid, 28049 Madrid, Spain \\
$^{7}$ Argelander-Institut f\"ur Astronomie, Auf dem H\"ugel 71, 53121 Bonn, Germany \\
$^{8}$ Leiden Observatory, Leiden University, Niels Bohrweg 2, 2333CA, Leiden, The Netherlands \\
$^{9}$ Department of Physics \& Astronomy, University College London, Gower Street, London WC1E 6BT, UK \\
$^{10}$ Institute for Computational Cosmology and Centre for Extragalactic Astronomy, Department of Physics, Durham University, \\ Durham DH1 3LE, UK  \\
$^{11}$ Instituci\'o Catalana de Recerca i Estudis Avan\c{c}ats (ICREA), 08010 Barcelona, Spain \\
}

\begin{document}

\maketitle

\begin{abstract}
The PAU Survey (PAUS) is an innovative photometric survey with 40 narrow bands
at the William Herschel Telescope (WHT). The narrow bands are spaced at 100\AA\
intervals covering the range 4500\AA\ to 8500\AA\ and, in combination with
standard broad bands, enable excellent redshift precision. This paper describes
the technique, galaxy templates and additional photometric calibration used to
determine early photometric redshifts from PAUS. Using \textsc{bcnz2}, a new
photometric redshift code developed for this purpose, we characterise the
photometric redshift performance using PAUS data on the COSMOS field.
Comparison to secure spectra from zCOSMOS DR3 shows that PAUS achieves
$\sigma_{68} /(1+z) = 0.0037$ to $i_{\mathrm{AB}} < 22.5$ for the redshift
range $0 < z < 1.2$, when selecting the best 50\% of the sources based on a
photometric redshift quality cut. Furthermore, a higher photo-z precision
($\sigma_{68}/(1+z) \sim 0.001$) is obtained for a bright and high quality
selection, which is driven by the identification of emission lines. We conclude
that PAUS meets its design goals, opening up a hitherto uncharted regime of
deep, wide, and dense galaxy survey with precise redshifts that will provide
unique insights into the formation, evolution and clustering of galaxies, as
well as their intrinsic alignments.
\end{abstract}


\section{Introduction}
Wide-field galaxy surveys are critically important when studying the late-time
universe. By mapping the positions, redshifts and shapes of galaxies, we are
able to measure the statistical properties of the cosmological large-scale
structure, which in turn allows us to make inferences on, for instance, the
nature of dark energy and dark matter \citep{Weinberg2013}. In cosmology, these
wide-field surveys are typically divided into two types: spectroscopic surveys
and imaging surveys.

Deep spectroscopic redshift surveys typically cover relatively small areas, but
with a high galaxy density \citep[e.g.][]{Davis2003, Lilly2007}. Such
observations have shown how the physical properties of galaxies depend on their
environment and how these evolve over time \citep[e.g.][]{Tanaka2004}. Such
targeted studies, however, are limited to relatively small physical scales. In
contrast, surveys probing large scales only sparsely sample the density field
\citep[e.g.][]{Strauss2002}. This allows them to infer cosmological parameters
by mapping the spatial distribution of galaxies on large scales. Moreover, the
targets are typically preselected, to efficiently get redshifts with minimum
observation time \citep[e.g.][]{Jouvel2014}.

Complete spectroscopic redshift coverage of a large area is difficult with
current instrumentation. Multi-object fibre spectrographs on 4m class
telescopes have surveyed large areas of sky, but fibre collisions limit
the efficiency with which small scales can be probed. It is, however,
possible to achieve a high spatial completeness as demonstrated by the Galaxy
Mass Assembly (GAMA) survey \citep{Driver2009}. This project used the AAOmega spectrograph
on the Anglo-Australian Telescope (AAT) to obtain $\sim 300,000$ spectroscopic
redshifts down to $r<19.8$ mag over an area of almost 300 deg$^2$. Repeated
observations allowed a 98\% completeness down to the limiting magnitude. The
bright limiting magnitude, however, limits the analysis to relatively low
redshifts and relatively luminous galaxies. Large telescopes are needed to
probe higher redshifts, but their field-of-view is typically too small to cover
large areas.

As a consequence, the role of environment on intermediate to small scales (below
10-20 Mpc), i.e. the weakly non-linear regime, is not well studied.
Interestingly this is where the statistical signal-to-noise is highest for
large galaxy imaging surveys. To robustly separate cosmological and galaxy
formation effects, we need to dramatically improve our understanding of these
scales, where baryonic and environmental effects become relevant. This requires
surveying large contiguous areas while simultaneously achieving a high density
of galaxies with sub-percent photometric redshift accuracy.
In this paper we present the first results of an alternative approach that
enables us to survey large areas efficiently, whilst achieving excellent
redshift precision for galaxies as faint as $i_{\mathrm{AB}}\sim 22.5$. 

The Physics of the Accelerating Universe Survey (PAUS) at the William Herschel
Telescope (WHT) uses the PAU Camera \citep[PAUCam,][]{PAUcam} to image the sky with 40 narrow
bands (NB) that cover the wavelength range from 4500\AA\ to 8500\AA\ at
100\AA\ intervals. These images are combined with existing deep broad band (BB)
photometry. Based on simulations \citep{Marti2014}, the expected photo-z
precision is $\sigma_{68} / (1+z) = 0.0035$ for $i < 22.5$ for a 50\% quality
cut. The quality cut is based on the posterior distribution and does not
use spectroscopic information. This precision corresponds to $\simeq 12$
Mpc/$h$ in comoving radial distance at $z=0.5$\footnote{Throughout the paper we
use a Planck2015 \citep{Planck2015} cosmology with $h = 0.68$.}. The initial motivation to reach
such precision was to be able to resolve the baryon acoustic oscillations (BAO)
peak \citep{Benitez2009, Gaztanaga2009}, but it also allows us to probe the start of the
weakly non-linear regime for structure formation. Moreover, this precision is
(nearly) optimal for many cosmological
applications \citep{Gaztanaga2012,Eriksen2015}. 

Cosmological redshifts are traditionally determined either from spectra or
broad band photometry. The redshift precision that can be achieved using
broad band photometry is typically $\sigma_{68}/ (1+z)  \simeq 0.05$
\citep[e.g.][]{Hildebrandt2012,Hoyle2018}, while including infrared, ultra violet (UV) and
intermediate bands can reduce the uncertainties by factors of a few
\citep{Laigle2016,Molino2014}. The much higher wavelength resolution of
spectrographs allows for a much improved determination of the locations of
spectral features, resulting in  high precision redshifts $\sigma_{68}/ (1+z)
\lesssim 0.001$. Many applications, however, do not require such precision and the
predicted PAUS performance is more than adequate.

For instance, errors in photo-z estimates translate into errors in the
luminosity or star formation rate (SFR). At $z=0.5$ the typical broad band photo-z
uncertainty of $\sigma_{68}/ (1+z)  \simeq 0.05$  translates into a 40\% error
in the luminosity (or 355 Mpc/$h$ in luminosity distance), while the PAUS
photo-z error corresponds to 2.5\%, comparable to other sources of errors (such
as flux calibration). For clustering measurements the improvement is even more
important as the uncertainty in comoving radial distance is reduced by more
than an order of magnitude from 171 Mpc to 12 Mpc, sufficient to trace the
large-scale structure. The improvement provided by spectroscopic redshifts,
which are typically ten times better, is therefore of limited use. 

Even though PAUS will cover a modest area compared to large wide imaging
surveys, PAUS will increase the number density of galaxies with sub-percent
precision redshifts by nearly two orders of magnitude to tens of thousands of
redshifts per square degree. Such redshift precision over a large area will
allow a range of interesting studies. It enables the study of the clustering of
galaxies in the transition from the linear to non-linear regime with high
density sampling for several galaxy populations. This will also allow multiple
tracer techniques over the same dark matter field \citep{Eriksen2015, Alarcon2018}. 

An important application is the study of the intrinsic alignments of galaxies.
These are an important tracer of the interactions between the cosmic
large-scale structure and galaxy evolution processes
\citep[e.g.][]{Catelan2000, Heavens2000, Croft2000, Hirata2004,
Joachimi2015,Troxel2015}. They are also a limiting astrophysical systematic in cosmic
weak lensing surveys, especially for the next generation of dark energy
missions, such as LSST \citep{lsst1}, {\it Euclid} \citep{euclidrb} and
{\it WFIRST} \citep{wfirst}. The depth of the PAUS data will push the
measurements out to $z\sim 0.75$, allowing us to study the luminosity and
redshift dependence of the signal, whilst at the same time probing a wide range
of environments. By targeting fields for which high-quality shape measurements
already exist (CFHTLS W1, W2, W3 and W4), PAUS is expected to achieve
competitive intrinsic alignment measurements.

In this paper we present the first results for PAUS, demonstrating
that we can indeed achieve the predicted redshift precision. The analysis in
this paper is limited to PAUS observations of the COSMOS
field\footnote{\url{http://cosmos.astro.caltech.edu/}}. This is a well-studied
area on the sky with a wide range of ancillary data, such as high-resolution HST
imaging, and deep broad- and medium-band imaging data extending both towards UV and
near infrared (NIR) wavelengths. Importantly for this study, extensive spectroscopy is
available. This enables us to quantify the precision with which we can determine
redshifts and compare the results to the predictions based on simulated data.

The structure of the paper is as follows. In \S \ref{data} we give an overview of
the PAUS data reduction and external data used. In \S \ref{pau_data} we present the
PAUS data in the COSMOS fields and the PAUCam filters. We introduce the
\textsc{bcnz2} code in \S \ref{photoz_code} and give additional details in
Appendix \ref{app_photoz_code}. Sections \S \ref{photoz_results} and \S
\ref{photoz_trends} details the photo-z results. Additional background material
and results can be found in Appendices \ref{add_results} and
\ref{dust_extinction}.

\section{Data}
\label{data}
In \S\ref{sec:data_overview} we briefly discuss the PAUS data reduction, while
\S \ref{sec:externalBB} presents the external broad band data. The
spectroscopic redshift catalogue to validate the photo-z performance is
described in \S\ref{sec:spectroscopy}.

\subsection{Data reduction overview}
\label{sec:data_overview}

To efficiently process the large amount of data from PAUS, a dedicated data
management, reduction, and analysis pipeline has been developed (PAUdm). We refer the
interested reader to the specific papers that describe the various steps in
more detail, including the associated quality control.

Following the observations, the raw data are transferred and stored at Port
d'Informaci\'{o} Cient\'{i}fica (PIC) \citep{PAUdm}. The day after the data
are taken, the images are processed there, using the \textsc{nightly} pipeline
\citep{PAUimage,PAUcalib}. This pipeline performs basic instrumental
de-trending processing, some specific scattered light correction and finally an
astrometric and photometric calibration of the narrow band images.

The master bias is constructed from  exposures, with a closed shutter and zero
exposure time, using the median of at least five images. Images are then flattened
using dome flats, obtained by imaging a uniformly illuminated screen.
Cosmic rays are removed with Laplacian edge detection \citep{Dokkum2001}. A
final mask also removes the saturated pixels.

In order to properly align the multiple exposures an astrometric solution is
added. We use the \textsc{astromatic} software\footnote{\url{https://www.astromatic.net/}}
\citep[\textsc{sextractor}, \textsc{scamp}, \textsc{psfex}][]{Bertin2011}. An
initial catalogue is created using \textsc{sextractor} \citep{Bertin1996}. The
astrometric solution is then found using \textsc{scamp} by comparing to {\it Gaia}
DR1\citep{Brown2016}. Furthermore, the point spread function (PSF) is modelled
with \textsc{psfex}. Stars in the COSMOS field are identified through
point-sources in the COSMOS-Advanced Camera for Surveys (ACS)
\citep{Koekemoer2007,Leauthaud2007}, as available from \citet{Laigle2016}
(hereafter COSMOS2015). For the wide fields, we have developed a new method,
separating stars and galaxies with convolutional neural
networks (CNN) using the the narrow band data \citep{Cabayol2018}.

PAUS is calibrated relative to the Sloan Digital Sky Survey (SDSS) \citep{PAUcalib}.
Stars in the overlapping area with $i < 21$ are fitted with the Pickles stellar
templates \citep{Pickles1998} using the SDSS \bi{u},\bi{g},\bi{r},\bi{i} and
\bi{z} bands \citep{Smith2002}. The corresponding spectral energy distribution
(SED) and best fit amplitude then provide a model flux in the narrow bands. To
ensure a robust solution, we limit the calibration to stars with
$\mathrm{SNR}>10$ in the narrow band and $i_{\mathrm{AB}} < 21$. A single
zero-point per image is determined by comparing the model and observed fluxes.
The calibration step removes the Milky Way (MW) extinction. When fitting to
SDSS, the model includes extinction. For the correction we use the
corresponding model without extinction.

The galaxy fluxes are measured by the \textsc{memba} pipeline \citep{PAUphoto}.
Deeper broad band (BB) data exist for both COSMOS and the wide fields. Hence the galaxy
positions are determined a priori using these data and the narrow-band (NB)
fluxes are determined using forced photometry by placing a suitable aperture on
the NB images, centred on these positions. To provide consistent colours, we
match the aperture to the size of the galaxy of interest, using the r50 deconvolved
measurement from COSMOS ACS. In the case of the COSMOS data the size
and elliptical shape used comes from the COSMOS Zurich catalogue
\citep{Sargent2007}. The elliptical aperture in \textsc{memba} is scaled using both
the size and PSF FWHM to target 62.5\% of the total flux. While the
optimal SNR depends on the galaxy light profile and Sersic index, this fraction
is close to optimal.

Fluxes are measured on individual exposures, where the background is determined
using an annulus from 30 to 45 pixels around the galaxies. The galaxies falling
into the background annulus are removed with a sigma-clipping. The fluxes are
thus background subtracted, scaled with the image zero-points and then combined
with a weighted mean into coadded fluxes.

\subsection{External broad bands}
\label{sec:externalBB}

We used the BB data from the COSMOS2015 catalogue. It includes $u^*$ band data
from the Canada-France Hawaii Telescope (CHFT/MegaCam) and \bi{B}, \bi{V},
 \bi{r}, $i^+$, $z^{++}$ broad band data from Subaru, obtained as part
of the COSMOS-20 survey \citep{Taniguchi2015}. Figure \ref{bb_filters} shows
the broad band transmission curves. We use the 3'' diameter PSF homogenised
flux measurements available in the catalogue
release\footnote{\url{ftp://ftp.iap.fr/pub/from_users/hjmcc/COSMOS2015/}} and
apply several corrections as described and provided in COSMOS2015. 

The Milky Way interstellar dust reddens the observed spectrum of background
galaxies. As described in the previous subsection, PAUS data are corrected for
dust extinction in the calibration. Therefore we need to do the same for COSMOS
data. Each galaxy has an $E(B-V)$ value from a dust map \citep{Schlegel1998},
and \citet{Laigle2016} provide an effective factor $F_x$ for each filter
$x$ according to \citet{Allen1976}. For each galaxy the corrected magnitudes
are

\be
\text{Mag corrected}_x = \text{Mag uncorrected}_x - E(B-V) * F_x.
\ee

Photometric offsets are added to acquire total fluxes as described in
\citet{Laigle2016}. This is not strictly needed since the photometric code
estimates a zero-point shift between the broad and narrow band systems per
galaxy (see \S \ref{photoz_formalism}).

\subsection{Spectroscopic catalogue}
\label{sec:spectroscopy}
To determine the accuracy of the photometric redshift estimation using PAUS, we
compare to zCOSMOS DR3 bright spectroscopic data, which has a pure magnitude
selection in the range $15<i_{\mathrm{AB}}<22.5$ \citep{Lilly2007}.  This
selection yields a sample mainly covering the redshift range $0.1 \lesssim z
\lesssim 1.2$ in 1.7 deg$^2$ of the COSMOS field \citep[$149.47\degree \lesssim
\alpha \lesssim 150.77\degree$, $1.62\degree \lesssim \delta \lesssim
2.83\degree$][]{Knobel2012}.

This dataset contains 16885 objects of which 10801 remain after removing less
reliable redshifts based on a provided confidence class \citep[$3 \leq
\mathrm{CLASS} < 5$][]{Lilly2009}. This sample covers most of the redshift and
magnitude range for PAUS, which makes it especially interesting for validating
the photometric redshift precision. The spectroscopic completeness is shown in
Figure \ref{fig:cosmos_completeness}.

\section{PAUCam data in the COSMOS field}
\label{pau_data}

As the start of the survey suffered from adverse weather conditions, the data
for the COSMOS field were collected over a longer period in the semesters
2015B, 2016A, 2016B and 2017B. As detailed in \citet{Madrid2010, Castander2012,
PAUcam}, the narrow-band filters are distributed through 5 interchangeable
trays, each carrying a group of 8 NB filters consecutive in wavelength. Each
position is imaged with exposure times of 70, 80, 90, 110 and 130 seconds, from
the bluest to the reddest tray. The COSMOS field was divided into 390
pointings, each observed with between 3 and 5 dithers for each of the 5
narrow-band filter trays. The final data set, which lacks some pointings,
comprises a total of 9715 exposures.

\subsection{Filter transmission curves}
\label{filter_curves}

\begin{figure}
\xinclude{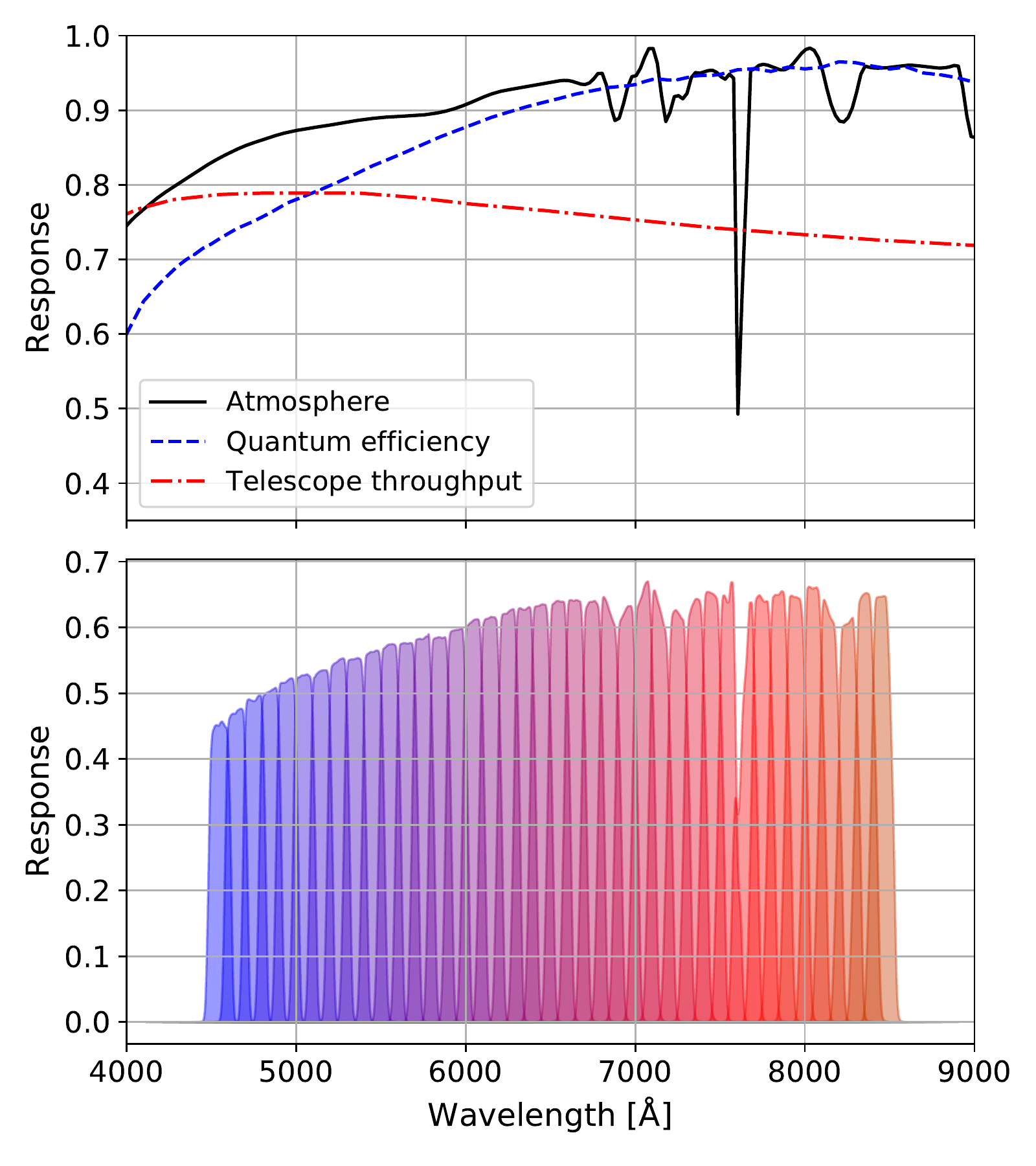}
\caption{\emph{Top:} The atmosphere, quantum efficiency and telescope
throughput. \emph{Bottom:} The throughput of the PAUS narrow bands when
combining the filter transmission and the effects in the top panel.}
\label{filters}
\end{figure}

The PAUCam \citep{PAUcam} instrument at the William Herschel Telescope (WHT) has a
novel set of 40 narrow band and 6 broad band filters. The total narrow band
transmission includes filters, atmosphere, instrument and telescope
effects. Figure \ref{filters} shows the filter transmission curves, where
the top panel shows the effect of the atmospheric transmission. As a
preliminary solution we used the Apache Point Observatory (APO) transmission
and will update this in due course. Any residual differences are removed in the
calibration step comparing with reference standard stars
\citep[see][]{PAUcalib}. The quantum efficiency (blue line) of the Hamamatsu
CCDs has been measured at the IFAE laboratories \citep{Casas2014}, while for
the telescope throughput (red line) we use the publicly available transmission
for the WHT.

In the bottom panel of Figure \ref{filters}, the narrow band throughput is
shown, including the effects mentioned above combined with filter
transmission. The optical filters, are 130\AA\ (FWHM) wide and equally spaced
(100\AA) in the range between 4500\AA\ and 8500\AA. The transmission was
measured in the CIEMAT optical laboratory and shifted to the PAUCam operating
temperatures using a theoretical relation \citep{Casas2016}.

\subsection{Signal to noise ratio}
\begin{figure}
\xinclude{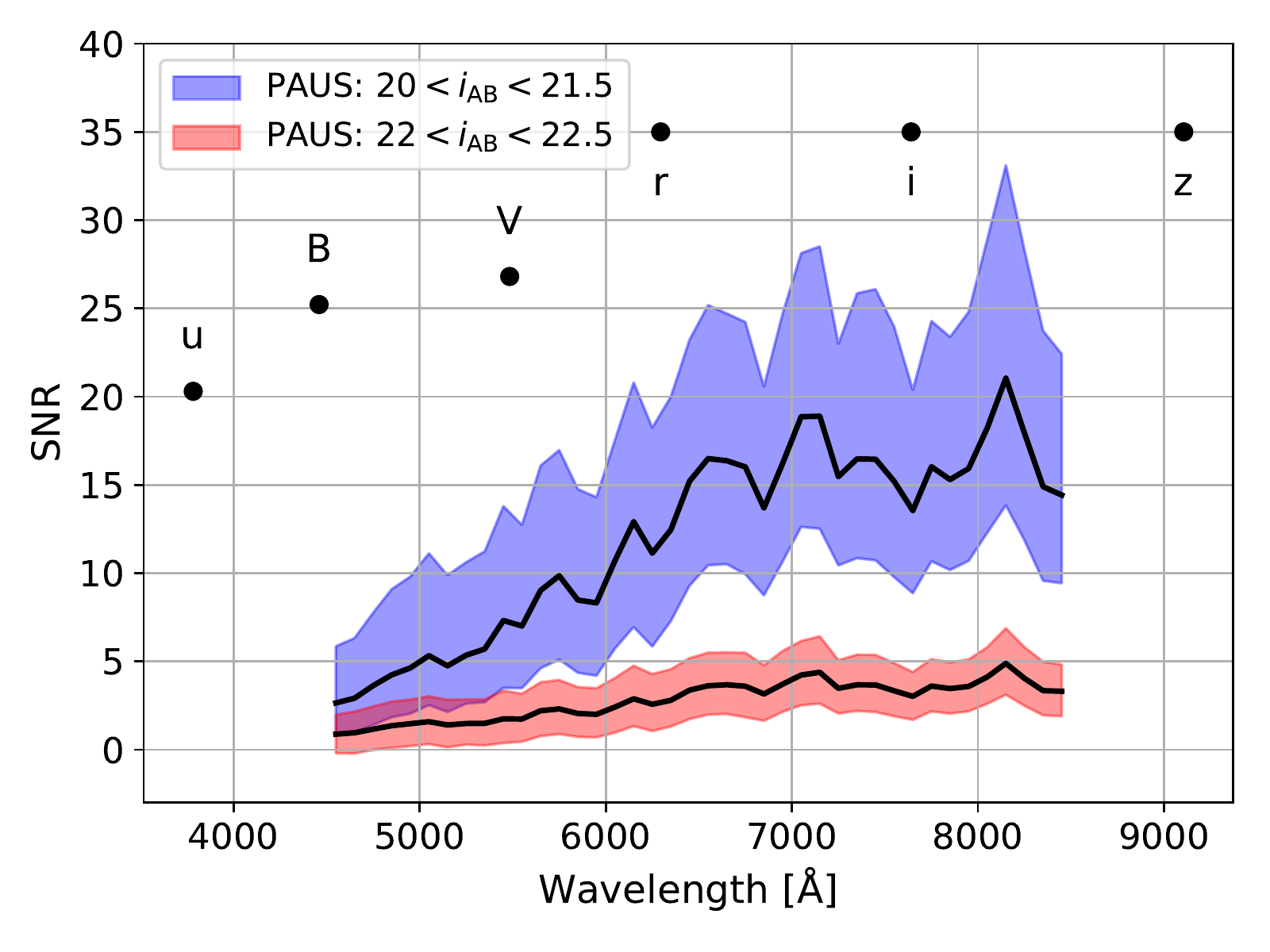}
\caption{The SNR per exposure distribution on the COSMOS field. Two lines shows
the median SNR in a bright and faint subsample. The surrounding shaded band
shows the area between 16 and 84 percentiles. For the broad bands, we only show
the median SNR of the faintest subsample.}
\label{pau_SN}
\end{figure}

The typical SNR (flux/error) per exposure of the narrow band flux measurements
is low, in particular at bluer wavelengths. The redder NBs are sky limited,
while the bluer bands are limited by the readout noise. For the wide fields
the exposure times were adapted to adjust for this lack of SNR in the bluer
bands. This is evident from Figure~\ref{pau_SN}, which shows the SNR for the
PAUCam data we use here. To illustrate the trends, the data are split into a
bright ($20<i_{\mathrm{AB}} < 21.5$) and faint ($22 < i_{\mathrm{AB}} < 22.5$)
subsample. The lines indicate the median SNR, while the filled bands show the
corresponding 16 and 84 percentiles.

In the bright subsample ($20<i_{\mathrm{AB}} < 21.5$), the median SNR increases
from 2.7 to 14.5 from the bluest (\bi{NB455}) to the reddest (\bi{NB855}) band.
Each tray contains 8 filters, so it is not possible to optimise the exposure
time for each filter. Moreover, most of the galaxies have red SEDs and thus are
brighter in the reddest bands.

The faint subsample ($22 < i_{\mathrm{AB}} < 22.5$) has a much lower SNR. As a
result, flux estimates can become negative due to noise. The median SNR in this
plot ranges from 0.9 to 3.3, from the bluest to the reddest band. It is
important that the photo-z codes properly handle the low SNR for individual NB
measurements, as they still contain information.

The black points in Figure \ref{pau_SN} indicate the median SNR measured in the
COSMOS2015 BB data for the faint sample. We limit the precision of these
measurement to 3\%  for all bands (see \S\ref{photoz_code}), i.e. limiting the
SNR to 35. This ensures that the broad band data do not dominate the fits as
the uncertainties for some of the BB data appear to be underestimated
\citep{Laigle2016}. The SNR of the BB data is about 8 times higher than with
the narrow bands, which can pose challenges for the photo-z determination. For
instance, it requires a careful calibration between the bands
(\S\ref{photoz_formalism}).

\section{Photometric redshift estimation}
\label{photoz_code}
Photometric redshift can be determined using a variety of approaches, and
consequently different public photo-z codes are available. Examples of
template-based codes include \textsc{bpz} \citep{Benitez2000} and
\textsc{lephare} \citep{Arnouts2011}. These compare the observations to
predefined redshift dependent models. Using machine learning the
\textsc{skynet} \citep{Bonnett2015}, \textsc{annz2} \citep{Sadeh2016}, and
\textsc{dnf} \citep{Vicente2016} codes can learn the relation between flux and
redshift.

While the public \textsc{bpz} code was used in \citet{Marti2014}, it does
not include emission lines in a flexible way (\S \ref{emission_lines}). These
lines are critical for achieving the required precision. This paper introduce
\textsc{bcnz2}, a new code specifically developed for the challenges found
using PAUS data.

\subsection{Model flux estimation}
\label{gen_model_flux}
The \textsc{bcnz2} is a template based photometric redshift code, that compares
the observed flux in multiple bands with redshift dependent models of the
galaxy flux. The observed flux is a wavelength dependent convolution of the
galaxy SED and the response of the detector. Let $f_{\lambda}(\lambda)$ be the
galaxy SED, which is the flux a galaxy transmits at a wavelength $\lambda$.
With the expansion of the Universe, a photon emitted at $\lambda_e$ is observed
at $\lambda_o = (1+z) \lambda_e$. The observed photon flux ($f_i$) in a fixed
band is \citep{Hogg2002,Marti2014}

\be
f_i = \int_0^{\infty} \mathrm{d}\lambda\, \lambda f_{\lambda}(\lambda) R_i(\lambda),
\ee
\noindent
where $R_i(\lambda)$ is the system response which is a multiplicative combination
of atmospheric, telescope, CCD detector and filter transmission (\S \ref{filter_curves}).
The galaxy SEDs $f_{\lambda}(\lambda)$ used are described in \S
\ref{combine_seds}.

\subsection{Photo-z formalism}
\label{photoz_formalism}
The \textsc{bcnz2} photo-z algorithm uses a linear combination of templates in
order to fit the measured fluxes. For each galaxy, we estimate the redshift
probability distribution $p(z)$ for a given galaxy defined as:

\newcommand{\fmodel}{f^{\mathrm{Model}}}

\be
p(z) \propto \underset{{\bm \alpha}_1 \geq 0}{\int \mathrm{d}{\bm \alpha}_1}
\dots \underset{{\bm \alpha}_n \geq 0}{\int \mathrm{d}{\bm \alpha}_n} \exp\left(-0.5
\chi^2[z, {\bm \alpha}]\right) p_{\mathrm{Prior}}(z, {\bm \alpha}) ,
\label{main_pz_eq}
\ee
\noindent
where $p_{\mathrm{Prior}}(z, {\bm \alpha})$ are the general form of the priors
and $n$ is the number of templates. Here, as described in \S
\ref{photoz_algorithm}, the integration is restricted to positive normalisation
of the templates (${\bm \alpha}_i$). Further, we define

\be
\chi^2[z, {\bm \alpha}] = \sum_{i,NB} \left(\frac{\tilde{f_i} - l_i k \fmodel_{i}}{\sigma_i} \right)^2
+ \sum_{i,BB} \left(\frac{\tilde{f_i} - l_i \fmodel_{i}}{\sigma_i} \right)^2,
\label{chi2_eq}
\ee

\noindent
where $l_i$ and $k$ are calibration factors, which are explained later.  
Here $\tilde{f}_i$ is the observed flux in band $i$, $\sigma_i$ is the 
corresponding error. The model flux, $\fmodel_i$, is defined by

\be
\fmodel_{i}[z, {\bm \alpha}] \equiv \sum_{j=1}^n f^j_i(z) {\bm \alpha}_j,
\ee

\noindent
where $f_i^j$ is the model flux of template $j$ in band $i$, with amplitude
${\bm \alpha}_j$. The final template is therefore a linear combination of
templates, which are defined in \S \ref{combine_seds}. This approach
is similar to that of  \textsc{eazy} \citep{Brammer2008}.

COSMOS photometry uses fixed apertures rescaled to total flux, while PAUS uses
matched apertures (see \S \ref{sec:data_overview}).  Furthermore, an
uncertainty in the flux fraction introduces an uncertainty when scaling to
total flux. To match the narrow and broad band systems, we consider the
scaling $k$ (\S \ref{subsub_k}) as a free parameter per galaxy.

In addition, Eq. \eqref{chi2_eq} contains a global zero-point $l_i$ for
each band $i$. The PAU survey relies on external observations for the
broad bands. This might mean different photometry, including different
aperture sizes. As described \S \ref{subsub_l}, we therefore want to
determine a zero-point correction ($l_i$) for each band. This will be the
same for all galaxies.

\subsection{Photo-z algorithm}
\subsubsection{P(z) approximation}

\label{photoz_algorithm}
Integrating over all amplitudes in Eq. \eqref{main_pz_eq} is numerically
expensive and makes us sensitive to the priors. While a closed-form solution
exists, this allows for negative amplitudes (${\bm \alpha}$). In practice allowing for
negative amplitudes introduces too much freedom, which degrades the redshift
precision. Allowing for negative amplitudes would e.g. lead to the OIII line
template fitting to spurious low flux measurements caused by negative
(inter-CCD) cross-talk.  Some of the positive amplitude combinations should
also be prevented, e.g.  through a more physical modelling of the SEDs, in
future work. We therefore approximate:

\begin{align}
p(z) &\propto \underset{{\bm \alpha}_1 \geq 0}{\int \mathrm{d}{\bm \alpha}_1} \dots
\underset{{\bm \alpha}_n \geq 0}{\int \mathrm{d}{\bm \alpha}_n} \exp\left(-0.5 \chi^2[z,
{\bm \alpha}]\right) p_{\mathrm{Prior}}(z, {\bm \alpha}) \\
&\approx \exp(-0.5 \chi^2_{\mathrm{Min}}[z]),
\label{pz_int}
\end{align}

\noindent
where the integral at each redshift is approximated using the maximum
likelihood conditional on $z$ (min $\chi^2$), with the proportionality constant
being determined by requiring that $p(z)$ integrates to unity. While this
approximation only uses the peak position, we find that this works sufficiently
well. 

\subsubsection{P(z) estimation (per galaxy)}
\label{pz_per_gal}
The minimum is determined using the algorithm of \citet{Sha2007}. This
algorithm ensures the amplitudes ${\bm \alpha}$ remain positive. It is also
proven to converge towards the global minimum of $\chi^2[z, {\bm \alpha}]$ for
a fixed redshift $z$. We therefore minimise the $\chi^2$ expression with
respect to the amplitudes (${\bm \alpha}$) on a redshift grid in the redshift
range $0.01 < z < 1.2$, using $\Delta z = 0.001$ wide redshift bins. For
further details see Appendix \ref{add_results}. 

\subsubsection{COSMOS/PAU calibration (per galaxy)}
\label{subsub_k}
The minimisation algorithm relies on the $\chi^2$ expression being on a
quadratic form. Extending to also determining $k$ (Eq. \ref{main_pz_eq}), the
galaxywise scaling between the narrow and broad band photometry is therefore
not straightforward. Instead, using the derivative of the $\chi^2$ relation
(Eq. \ref{main_pz_eq}) with respect to $k$, one can find the solution which
minimises the $\chi^2$ value. This gives the solution

\be
k = \frac{\sum_{i,\mathrm{NB}} \tilde{f}_i l_i \fmodel_{i} / \sigma_i^2}{\sum_{i, \mathrm{NB}} {(l_i \fmodel_{i})}^2/\sigma_i^2},
\ee

\noindent
where the sum over filters only includes the narrow bands. Also, to lower the
runtime, we only estimate the zero-point $k$ at every tenth step in the
iterative minimisation (of ${\bm \alpha}$), as described in \S \ref{pz_per_gal}.

\subsubsection{Zero point recalibration (per band)}
\label{subsub_l}
To determine the zero-points per band ($l$), a common approach is to compare
the photo-z code best fit model with the observed fluxes \citep{Benitez2000}.
This ratio can be used to determine a zero-point offset per band. To estimate
the bandwise zero-points, we only estimate the best fit model at the
spectroscopic redshift. This reduces the runtime by three orders of magnitude,
since one only has to evaluate the fit at one redshift per galaxy. After
determining the best fit model ($f^{\mathrm{Model}}$) by running the photo-z
code for a fixed spectroscopic redshift, one finds the zero-point in band $i$
by

\be
l_i = \mathrm{Median}[f^{\mathrm{Model}} / f^{\mathrm{Obs}}],
\label{determineli}
\ee

\noindent
where we use the median, instead of a weighted mean, because it reduces the
impact of outliers. When using spectroscopic redshifts, one should in theory
split into a training and validation sample. However, unlike e.g. machine
learning redshifts, we train one number per band and not per galaxy. The
zero-points are therefore less affected by overfitting. We have tested that
this does not significantly affect the results and we therefore do not split
the catalogue by default.

The photo-z code is first run 20 times at the spectroscopic redshift. At
the start the offsets per band, $l_i$, are assumed unity and they are updated
after each iteration using Eq. \eqref{determineli}. In this process the scaling
$k$ is kept free. Afterwards we run the photo-z using the final zero-points
($l_i$), also treating $k$ as a free parameter. 

\subsection{Combination of SEDs}
\label{combine_seds}
The basic formalism of using a linear combination of templates has a problem when
including intrinsic extinction (Appendix \ref{dust_extinction}). The dust extinction
is not an additional template, but a wavelength dependent effect that
multiplicatively changes the SEDs. The simplest solution is to generate new
SEDs for different extinction laws and extinction values ($E(B-V)$). These can
then directly be used in the photo-z code. While possible in theory, we find
that this gives too much freedom, reducing the photo-z performance.

Instead, we add priors to restrict the possible SED combinations. The
minimisation algorithm limits our choice of priors. We group together the
SEDs in different sets, discussed later in this subsection. Within these sets
the prior is unity, but zero outside. This can be used both to avoid combining
different $E(B-V)$ values and unphysical template combinations. Using this
prior, Eq. \eqref{main_pz_eq} reduces to

\begin{align}
p(z) &\propto \sum_{\mu} \int \mathrm{d}{\bm \alpha}^{\mu}_1 \dots \int
\mathrm{d}{\bm \alpha}^{\mu}_n \exp(-0.5 \chi^2[z, {\bm \alpha}^\mu]) \\ &\approx
\sum_{\mu} \exp\left(-0.5 \chi^2_{\mathrm{Min\, {\bm \alpha}^\mu}}[z]\right),
\label{photoz_approx}
\end{align}

\noindent
where the sum is over different sets of SEDs (${\bm \alpha}^{\mu}$) (which we
call runs), with the approximation being the same as in Eq. \eqref{pz_int}. In
practice, this means one can separately run the photo-z code for many different
SED combinations and then combine them later (Eq. \ref{photoz_approx}).

\begin{table*}
\begin{center}
\begin{tabular}{llll}
\toprule
Run \# &Lines & Ext law & SED \\
\midrule
1 & False & None & Ell1, Ell2, Ell3, Ell4, Ell5, Ell6 \\
2 & False & None & Ell6, Ell7, S0, Sa, Sb, Sc \\
3 & True & None & Sc, Sd, Sdm, SB0, SB1, SB2 \\
4 & True & None & SB2, SB3, SB4, SB5, SB6, SB7, SB8, SB9, SB10, SB11 \\
5 & False & None & \begin{minipage}{0.6\linewidth} BC03(0.008, 0.509), BC03(0.008, 8.0), BC03(0.02, 0.509), 
BC03(0.02, 2.1), BC03(0.02, 2.6), BC03(0.02, 3.75) \end{minipage} \\
6-15 & True & Calzetti & SB4, SB5, SB6, SB7, SB8, SB9, SB10, SB11 \\
16-25 & True & Calzetti+Bump 1 & SB4, SB5, SB6, SB7, SB8, SB9, SB10, SB11 \\
26-35 & True & Calzetti+Bump 2 & SB4, SB5, SB6, SB7, SB8, SB9, SB10, SB11 \\
\bottomrule
\end{tabular}
\end{center}
\caption{The configurations used for the photo-z code. In the first column is
the configuration number, while the second gives whether emission line templates
are added. A third column gives the extinction law, which is used for $E(B-V)$
values between 0.05 and 0.5, with 0.05 spacing. The SB templates are
from the BC03 library. In run \#5 we use six additional BC03 templates, with
their metallicity (Z) and age (Gyr) specified in parenthesis. When running with
the Calzetti law (\#6-35), we also include two variations with a 2175\AA\ bump
(see Appendix \ref{dust_extinction}).}
\label{sed_combinations}
\end{table*}

Table \ref{sed_combinations} describes the SED and extinction combinations
that are used when running the photo-z code. For the case of elliptical 
and red spiral galaxy templates (run \#1-2), we neither include emission lines
nor dust extinction. For starburst galaxies, as used in run \#3-4,
\citet{Ilbert2009} had problems reproducing the bluest colours in the
spectroscopic sample. Following that paper, we use 12 starburst galaxies
generated by the \citet{Bruzal2003} models\footnote{Available in the Lephare
source code.}. These have ages spanning from 3 Gyr to 0.03 Gyr. Combining run
\#1-2 and \#3-4 slightly decreases the photo-z performance.

Following \citet{Ilbert2013} and \citet{Laigle2016}, we include a new set of
BC03 templates (run \#5) assuming an exponentially declining SFR with a short
timescale $\tau= 0.3$ Gyr to account for a missing population of quiescent
galaxies. In addition, starburst templates are run using the Calzetti
extinction law \citep{Calzetti2000} and the two modified versions (Appendix
\ref{dust_extinction}) with $E(B-V)$ values between 0.05 and $0.5$ in 10 steps
(run \#6-35).

\subsection{Emission lines}
\label{emission_lines} \begin{table}
\begin{center}
\begin{tabular}{lrrr}
\toprule
{} &  $\lambda [\mathrm{\AA}]$ & Template 1 & Template 2 \\
\midrule
${\mathrm{H}}_{\alpha}$  &                      6563 &    1.77 &       - \\
${\mathrm{H}}_{\beta}$   &                      4861 &    0.61 &       - \\
${\mathrm{Ly}}_{\alpha}$ &                      1216 &       2 &       - \\
${\mathrm{NII}}_{1}$     &                      6548 &    0.19 &       - \\
${\mathrm{NII}}_{2}$     &                      6583 &    0.62 &       - \\
OII                      &                      3727 &       1 &       - \\
${\mathrm{OIII}}_{1}$    &                      4959 &       - &       1 \\
${\mathrm{OIII}}_{2}$    &                      5007 &       - &       3 \\
${\mathrm{SII}}_{1}$     &                      6716 &    0.35 &       - \\
${\mathrm{SII}}_{2}$     &                      6731 &    0.35 &       - \\
\bottomrule
\end{tabular}

\end{center}
\caption{Emission line ratios. In the second column is the central
wavelength. The third column contains the main emission line template,
with flux ratios relative to $\mathrm{OII}$. In the last column is
the OIII template, normalized relative to ${\mathrm{OIII}}_{1}$.}
\label{emission_line_table}
\end{table}

Table \ref{emission_line_table} contains the set of emission lines that
are used in this paper. The emission lines are parameterised using a set of
fixed amplitude flux-ratios. These are obtained from COSMOS2015
\citep{Laigle2016} and references therein. When estimating the fluxes, we
approximate the emission lines as a delta function. In this table, the fluxes
are normalised to the OII values. \citet{Beck2016} found comparable ratios.

The inclusion of emission lines can be done in different ways. One approach is to
add the emission lines as an additional separate SED. This can be thought of as
having a contribution from a very young stellar population. We have added the
emission lines using two templates, one that contains all emission lines in
Table \ref{emission_line_table}, except the OIII doublet, which is kept in
a separate template. This is needed to take into account the large variability
between OIII and $\text{H}_{\beta}$ lines. Running with a single emission line
template led to a significant degradation in the photo-z performance. So far
we have used common practice and not included BPT \citep{Baldwin1981}
information. Better modelling of emission lines is expected in future
developments.

\section{Results}
\label{photoz_results}
In this section we present the main photo-z results (\S \ref{scatter_outliers})
and the additional calibration (\S \ref{sec:zero_points}).  The benefits of
combining broad and narrow bands are discussed (\S \ref{bb_nb_comb}), before
describing priors (\S \ref{prior_info}) and quality cuts (\S
\ref{quality_cuts}). We validate the Probability Density Function (pdf) in \S
\ref{pdf_validate}.

\subsection{Photo-z scatter and outliers}
\label{scatter_outliers}

\begin{figure}
\begin{center}
\xinclude{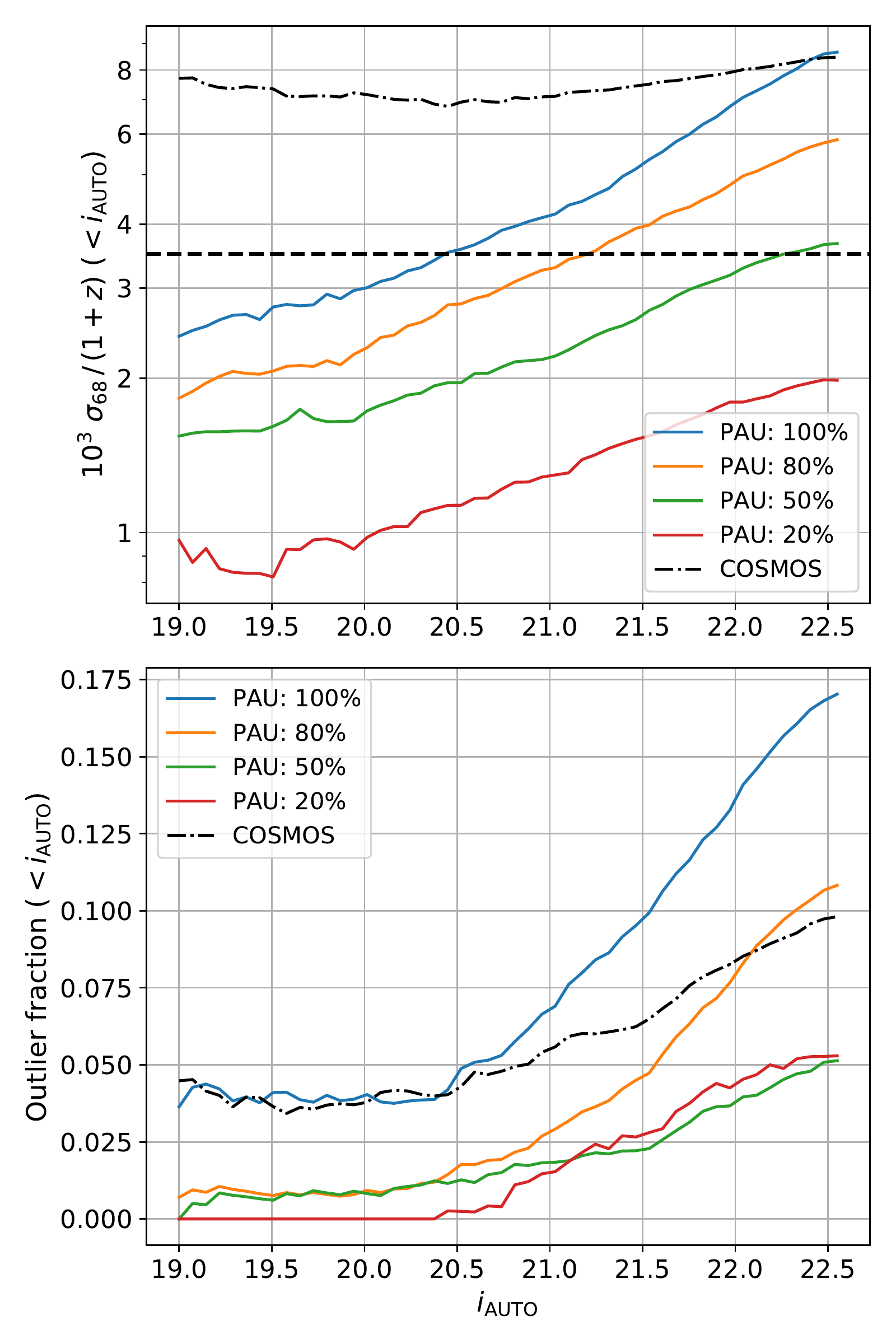}
\end{center}
\caption{The $\sigma_{68} / (1+z)$ (top) and outlier fraction (bottom) for
different quality cuts as a function of the cumulative magnitude bins. The
solid lines show the results when 100, 80, 50 and 20 percent of the sample
remain after a quality cut. The dashed line shows the COSMOS results without
any quality cuts, using the public COSMOS2015 catalogue.}
\label{main_photoz_cum}
\end{figure}

Figure \ref{main_photoz_cum} shows the main result of this paper: $\sigma_{68}$
and outlier fraction for PAUS and the COSMOS data. To quantify the photo-z
precision, we use

\be
\sigma_{68} \equiv 0.5\left(z^{84.1}_{\mathrm{quant}} - z^{15.9}_{\mathrm{quant}}\right)
\ee

\noindent
which equals the dispersion for a Gaussian distribution, but is less affected
by outliers. A galaxy is considered an outlier if 
\be
|z_\mathrm{p} - z_\mathrm{s}|\, /\, (1+z_\mathrm{s}) > 0.02, 
\ee

\noindent
where $z_\mathrm{p}$ and $z_\mathrm{s}$ are the photometric and spectroscopic
redshifts, respectively.

The COSMOS result uses the redshift estimate (zp\_gal) available in the
COSMOS2015 catalogue. The PAUS results are given for different fractions that
remain after a quality cut (Qz) (see \S \ref{quality_cuts}) based on PAUS
fluxes. Attempting to cut the COSMOS photo-z by the $p(z)$ quantiles
($z^{99}_{\mathrm{quant}}- z^{1}_{\mathrm{quant}}$) did not significantly
change their photo-z precision. We therefore only show the COSMOS
results for the full sample. The ALHAMBRA survey \citep{Moles2008} result is
not shown, since the public photo-z are worse than the COSMOS photo-z.

For $\sigma_{68}$ the horizontal lines marks the expected photo-z scatter
of $\sigma_{68} /(1+z) = 0.0035$ based on simulations at 50\% cut
\citep{Marti2014}. The PAUS photo-z is close to reaching this value, achieving
$\sigma_{68}/(1+z) \sim 0.0037$ for 50\% of the galaxies with $i_{\mathrm{AB}}
< 22.5$ and the spectroscopic selection shown in Figure
\ref{fig:cosmos_completeness}. Here the median $i_\mathrm{AUTO}$ is 20.6, 20.8,
21.2 and 21.4 for the 20, 50, 80 and 100 percent cuts, respectively. The
corresponding figure in differential magnitude bins and photo-z scatter plot
are included in Appendix \ref{add_results}. 

When applying a more stringent quality cut leaving less of the sample, the
$\sigma_{68}$ is approaching $0.001(1+z)$ for a bright selection and increases
to $0.002(1+z)$ for $i_{\mathrm{AB}} < 22.5$. While the selection on the
quality parameter results in selecting brighter galaxies, this population of
galaxies with quasi-spectroscopic redshifts was never seen in simulations
\citep{Marti2014}. Even when running \textsc{bpz} on a noiseless catalogue,
the $\sigma_{68}$ was never below $0.002(1+z)$. This mainly comes from emission
lines not being properly included in the simulations.

The bottom panel of Figure \ref{main_photoz_cum} shows the corresponding
outlier fraction. For the full sample, the PAUS photo-z has 18\% outliers for
$i_\mathrm{AB} <22.5$. This is higher than for COSMOS. Applying the quality cut
lowers the outlier rate to a more reasonable level. One should keep in mind
that the outlier rate is expected to reduce with better data reductions and
improvements to the photo-z code.

\subsection{Zero-points between systems}
\label{sec:zero_points}
\begin{figure}
\xfigure{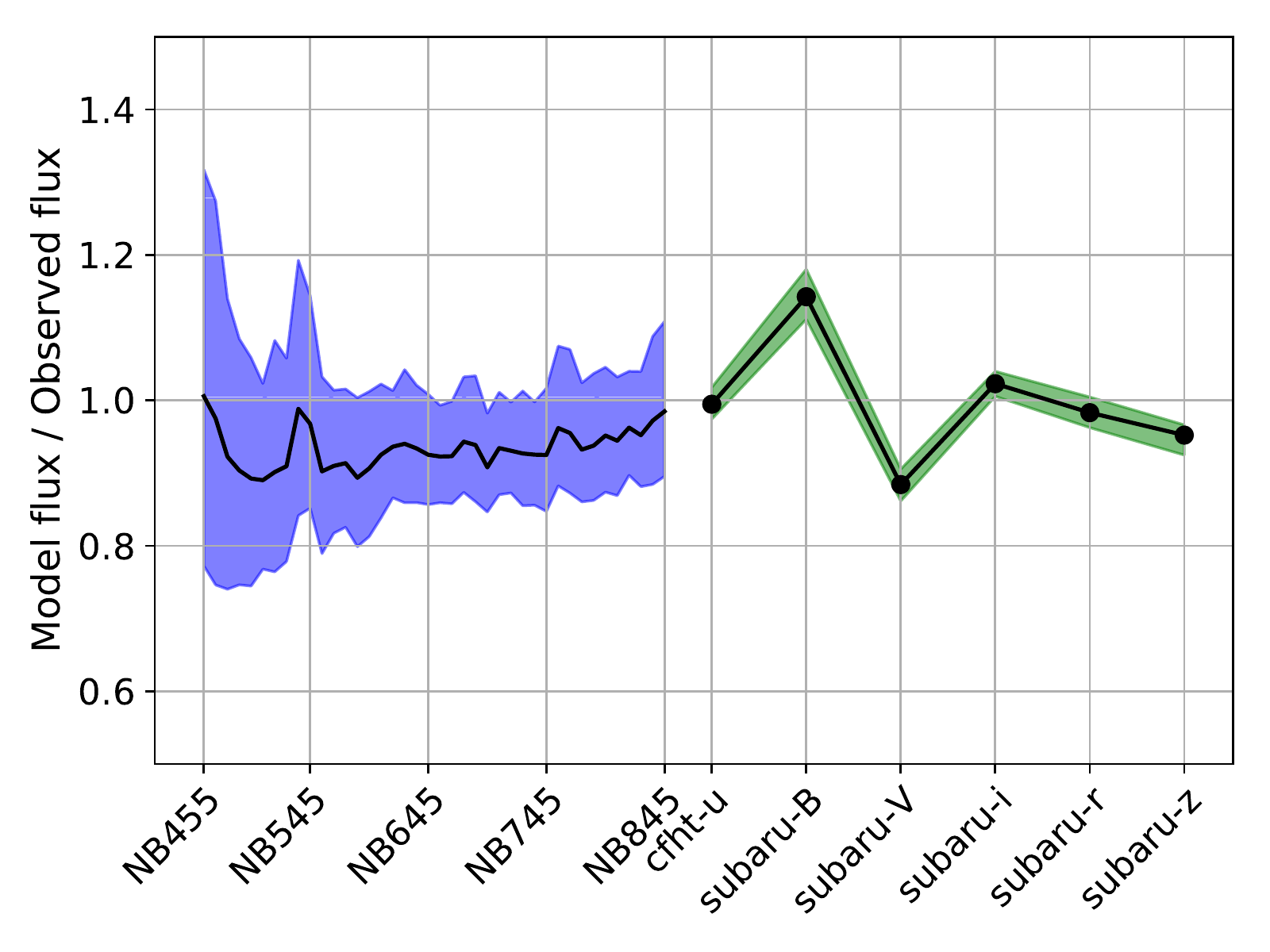}
\caption{The bandwise calibrations for the narrow (hatched) and broad (solid)
bands. On the x-axis is the band, while the y-axis shows the zero-points. The
solid line shows the median zero-point, while the bands show the 16 to 86
percentile interval.}
\label{calibration}
\end{figure}

Figure \ref{calibration} shows the recovered zero-points ($l_i$) from the
photo-z code. PAUS is already calibrated relative to SDSS stars, which have a
higher signal-to-noise ratio. One could restrict the additional zero-point
calibration to determining the BB zero-point from a model fit to the NB. In practice
we find better results from fitting to both NB and BB data, applying
zero-points to both systems. Including the BB decreases the model fit
uncertainty, but then includes bands which might require an offset. We handle
this by repeatedly estimating the best fit model and applying the resulting
offsets. By default this procedure is run with 20 iterations.

The x-axis shows the band, starting with the narrow band first and then the
uBVriz bands. The coloured band shows the region between 16 and 84 percentiles
of the offsets obtained from different galaxies in the last iteration step.
Here and in the final zero-points, we have only included measurements with $\mathrm{SNR}
> 1$. While the spread for individual galaxies is quite large, the mean value of 
the sample is centred around unity. For the narrow bands there is a tilt at the
blue end. When estimating the zero-points with only narrow bands (not shown),
the BB zero-points only change slightly.

\subsection{Combining broad and narrow bands}
\label{bb_nb_comb}
\begin{figure}
\xfigure{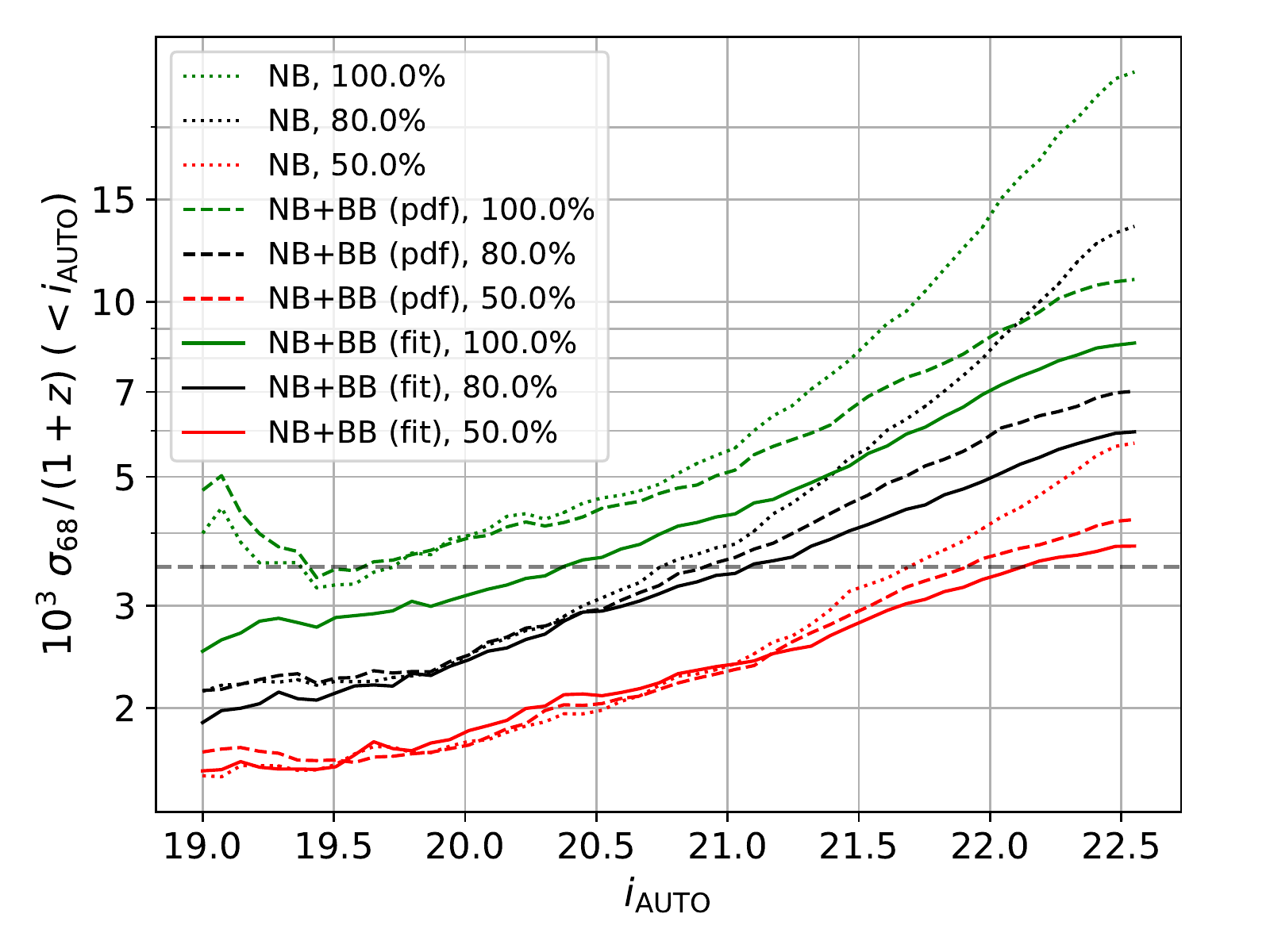}
\caption{The impact on photo-z precision of different approaches to combine NB
and BB information. The dotted lines show the narrow band performance alone,
while dashed lines (pdf) combine the NB and BB pdfs. The solid lines (fit)
simultaneously fit the narrow and broad bands.}
\label{nb_bb_comb}
\end{figure}

While the narrow bands are important, the broad bands also contribute
to photo-z precision. The broad bands have higher SNR (Fig. \ref{pau_SN}) and
cover a larger wavelength range (Fig. \ref{filters}). Qualitatively, these
determine the best fit SED and a broad redshift distribution, which acts as a
prior for the narrow bands. The narrow bands with good spectral resolution then
determine the redshift more precisely. Without the broad bands, the photo-z
code ended up confusing different emission lines. In particular, it confused
OIII and $\mathrm{H_\alpha}$, which led to redshift outliers with
$(z_\mathrm{p} - z_\mathrm{s}) / (1 + z_\mathrm{s}) \approx \pm 0.15$, with
more galaxies being scattered to lower redshift. Adding the broad bands
effectively solves this problem.

Figure \ref{nb_bb_comb} compares different ways of including the broad band
information in the photo-z code. The dotted lines show $\sigma_{68} / (1+z)$
when using narrow bands only. When running with NB alone, we combine the two
emission line templates. Then the combination with broad bands is done in
two different ways. First, we estimate the photo-z independently for the
narrow and broad bands. These are then combined by multiplying the pdfs

\be
p(z) = p_{\mathrm{NB}}(z)\times p_{\mathrm{BB}}(z),
\ee

\noindent
which is only approximately correct, since we have marginalized over the SEDs
independently for both runs. When adding the broad bands there is a significant
improvement in photo-z performance for all selection fractions. The correct and
more optimal approach is to estimate the photo-z, including both the broad and
narrow bands. This jointly constrains both the redshift and SED combination
from both systems, leading to a further decrease in the photo-z scatter.
Fitting NB+BB is better than combining pdfs of separate NB and BB fits. In 
Figure \ref{nb_bb_comb} the 20\% lines were removed, since they looked similar
for all methods. 

\subsection{Photo-z priors}
\label{prior_info}
\begin{table}
\begin{center}
\begin{tabular}{lrrr}
\toprule
{} &  No priors &  SED priors &  SED,z priors \\
Fraction &            &             &               \\
\midrule
100\%     &        8.5 &         8.5 &           8.3 \\
80\%      &        6.2 &         6.1 &           5.9 \\
50\%      &        3.9 &         3.9 &           3.7 \\
20\%      &        2.2 &         2.1 &           2.1 \\
\bottomrule
\end{tabular}

\caption{The ${10}^3\ \sigma_{68} / (1+z)$ values for different priors. The
first column gives the fraction of galaxies remaining after a quality cut (Qz),
while the second is the result without priors. In the third column the priors
are only applied to the SED combinations, while a fourth column adds priors 
(independently) on both the SED and redshift.}
\label{prior_table}
\end{center}
\end{table}

Template-based photometric redshift codes estimate the redshift by comparing
the observations to model fluxes, estimated by redshifting templates.
Estimating the redshift distribution requires, if using Bayesian methodology,
the inclusion of priors. These can significantly improve the redshift
estimation. When observing galaxies in a few colours or a restricted wavelength
range, some low and high redshift models have similar colours. A prior based on
luminosity functions effectively determines which solution is most probable
\citep{Benitez2000}. In this paper we include priors on redshift and SEDs, but
not on luminosity.

The PAU Survey observes galaxies with 40 narrow bands and combines these with
traditional broad bands. In addition, the PAU Survey mostly observes galaxies
in the redshift range $0 < z < 1.2$. Our redshift estimates should therefore be
less sensitive to colour degeneracies. However, we attempt to further improve
the redshifts by adding priors, constructed from the ensemble of galaxies.

The algorithm used when estimating the photometric redshift relies on the
$\chi^2$ expression to be quadratic in the model amplitudes (see \S
\ref{photoz_code}). This would make adding priors on the detailed SED
combinations difficult. However, we can add priors on the different photo-z
runs (Table \ref{sed_combinations}). This effectively adds priors on
the galaxy SED, the extinction law and the $E(B-V)$ value.

Table \ref{prior_table} compares the photo-z scatter for different priors
(columns) and fractions remaining after a quality cut (first column). The priors
for individual galaxies are constructed from the ensemble of galaxies. After
running the photo-z code once, we construct the priors combining the
probability for all galaxies. The second column gives $\sigma_{68} / (1+z)$
without priors, while the third column adds priors on each photo-z run. These
are obtained by first running the photo-z without priors and then construct
priors by the amount of galaxies having a minimum $\chi^2$ corresponding to
each of the photo-z runs. This gives a minor improvement for 100 percent of the
sample. 

Similarly, the last column combines priors on SEDs and the redshift
distribution. The optimal approach is to construct priors on both SEDs and
redshifts combined, but this led to a too noisy distribution, for too few
galaxies. Instead we combine the previous SED priors with a redshift prior as
independent priors. The redshift priors are constructed from the redshift
distribution obtained without a prior, convolved with a $\sigma_\mathrm{z} =
0.003$ Gaussian filter to smooth the distribution. The final priors improve
the photo-z for all selection fractions. This effectively also incorporates
some clustering information from the field.

\subsection{Quality cuts}
\label{quality_cuts}

\begin{figure}
\begin{center}
\xinclude{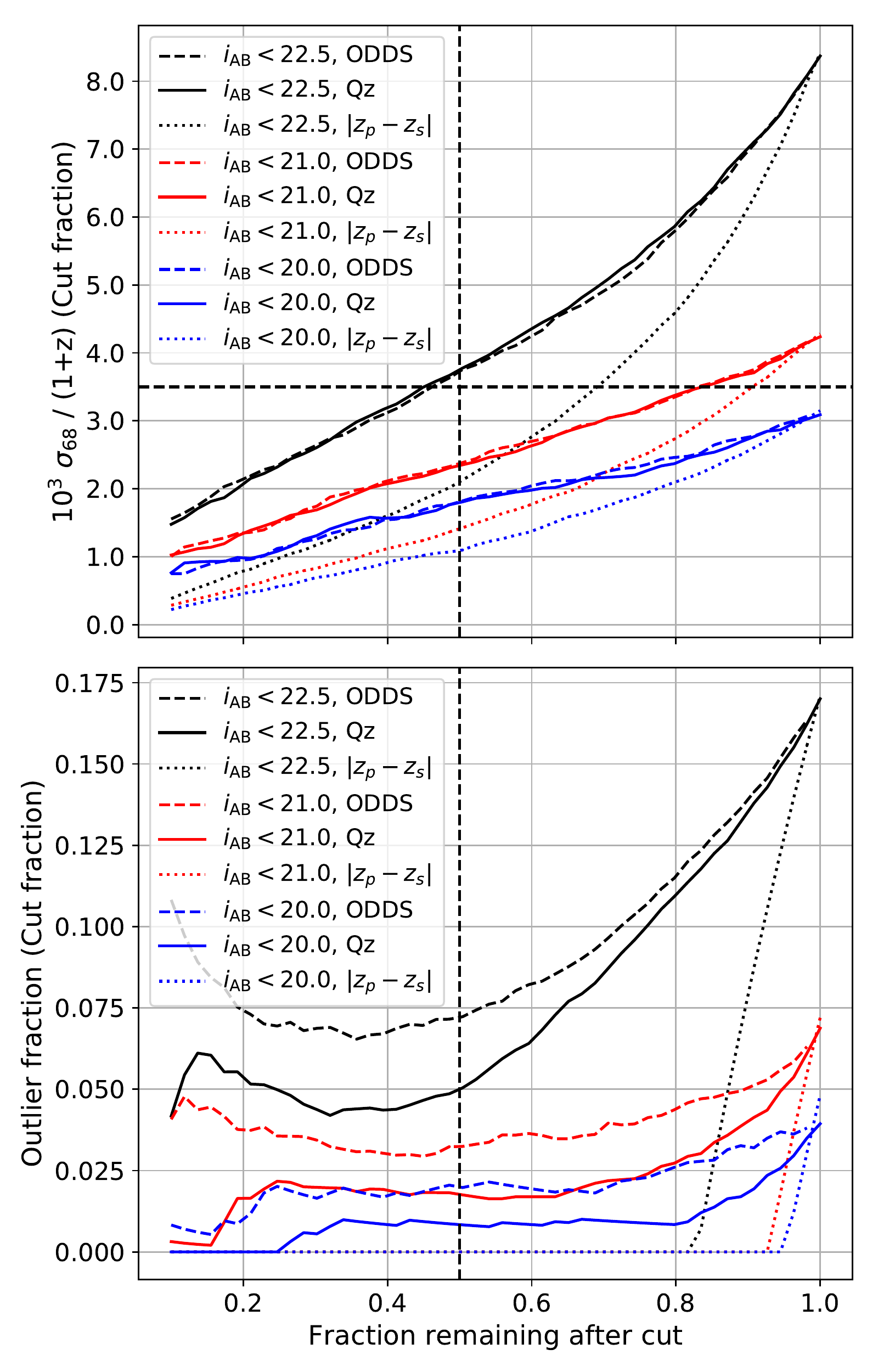}
\end{center}
\caption{The $\sigma_{68}$ (top) and outlier fraction (bottom) for different
magnitude limited samples and quality cuts as a function of the cut fraction.
The results are shown for the three magnitude cuts: $i_{\mathrm{AB}} < 20, 21,
22.5$ and two quality estimators: ODDS, Qz. Here $z_p - z_s$, which cuts on
the absolute different between the photometric and spectroscopic redshift,
is included as a reference. The horizontal dashed line (top panel) shows the
nominal PAUS photo-z precision target for a 50\% quality cut.}
\label{fig_quality_cuts}
\end{figure}

For different purposes, one might want to select a subsample with better
photo-z precision \citep{ElvinPoole2017}. A frequently used photo-z quality
parameter is the ODDS parameter \citep{Benitez2000} (\textsc{bpz}). The ODDS is
defined as 

\be
\mathrm{ODDS} \equiv \int_{\mathrm{z_b-\Delta z}}^{\mathrm{z_b+\Delta
z}}\mathrm{d}z\ p(z),
\ee

\noindent
where $z_b$ is the posterior redshift mode (peak in $p(z)$) and $\mathrm{\Delta
z}$ defines an interval around the peak, typically related to the photo-z
scatter.  This definition measures the fraction of the $p(z)$ located around
the redshift peak, which e.g. can be used to remove galaxies with double peaked
distributions. In this paper we use $\mathrm{\Delta z}=0.0035$, which is
reduced from typical broad band values since the PAUS pdfs are narrower.

One should be aware that such a selection can introduce
inhomogeneities. The photometric redshift quality flags depend on the
data quality, the galaxy SED, the modelling and the photo-z method. A
selection with a photo-z quality cut can indirectly cut on any or all
of these quantities. As an example, \citet{Marti2014b} found that
cutting on ODDS resulted in a spatial pattern corresponding to scanning
stripes in SDSS data.

The ODDS quality parameter contains information on the redshift
uncertainty, as described by the posterior $p(z)$. However it does not
give the goodness of fit. An alternative approach is to directly cut on
the $\chi^2$ from the fit (Eq. \ref{pz_int}). Removing galaxies
with a high $\chi^2$ improves the photo-z performance of the ensemble.
However, cutting on ODDS directly is more effective. Applying first a
$\chi^2$ cut and then an ODDS cut, always removing the same number of
galaxies, showed a better result than cutting only based on the ODDS.

Another photo-z quality parameter is Qz \citep{Brammer2008}, which
attempts to combine various quality parameters in a non-linear manner.
It is defined by

\be
\mathrm{Qz} \equiv \frac{\chi^2}{N_\mathrm{f}-3}\left(
 \frac{z^{99}_{\mathrm{quant}} - z^{1}_{\mathrm{quant}}}
{\mathrm{ODDS}(\mathrm{\Delta z}=0.01)}\right),
\label{qz_def}
\ee

\noindent
where $N_\mathrm{f}$ is the number of filters and $\chi^2$ is from the template
fit. The $z^{99}_{\mathrm{quant}}$ and $z^{1}_{\mathrm{quant}}$ are the 99 and
1 percentiles of the posterior distribution, respectively. The value $\Delta z
= 0.01$ in the ODDS, is adapted to match the narrower pdfs in PAUS.

Figure \ref{fig_quality_cuts} shows the $\sigma_{68}/(1+z)$ (top) and the
outlier fraction (bottom) for different magnitude cuts. Note, this interval is
about an order of magnitude smaller than what is typically used for broad band
photo-z estimates. Selecting the 50\% of the galaxied based on ODDS or the Qz
quality parameter both gives a $\sigma_{68}/(1+z)$ around $0.004(1+z)$ for
$i_{\mathrm{AB}} < 22.5$. As a reference, we have included the $|z_\mathrm{p} -
z_\mathrm{s}|$ line, which is the result when directly cutting on the absolute
difference to the spectroscopic redshift. Cutting on $|z_\mathrm{p} -
z_\mathrm{s}|$ is the best quality cut possible. In that case, the photo-z
scatter would be $0.0022(1+z)$ at 50\% and $i_{\mathrm{AB}} < 22.5$. The
performance therefore has some further room for improvement.

For $i_{\mathrm{AB}} < 22.5$ the outlier fraction (bottom panel) is
6.3\% when cutting 50\% of the galaxies with the Qz parameter. This is
lower than when selecting on ODDS, which in comparison has 7.8\%
outliers. In addition, the Qz parameter performs better when selecting
a lower fraction of galaxies and for a brighter sample. By default we
therefore use the Qz quality parameter throughout the paper.

Lastly, the $|z_\mathrm{s} - z_\mathrm{p}|$ (dotted) lines contain information
on the outlier fraction. The outlier fraction is 19.8, 9.3 and 5.9 percent at
$i_{\mathrm{AB}} < 22.5,\ 21\ \mathrm{and}\ 20$, respectively.  Attempting to
further reduce the outlier fraction will be an important part of future photo-z
developments.

\subsection{Validating the pdfs}
\label{pdf_validate}
The \textsc{bcnz2} code produces a redshift probability distribution
for each galaxy. Most results throughout this paper use the mode of the
distribution. For some science cases, one might want to weight based on the
redshift probability distribution \citep{Asorey2016}. A misestimation of the
pdf can then end up biasing the final quantity \citep{Nakajima2012}.

\begin{figure}
\begin{center}
\xinclude{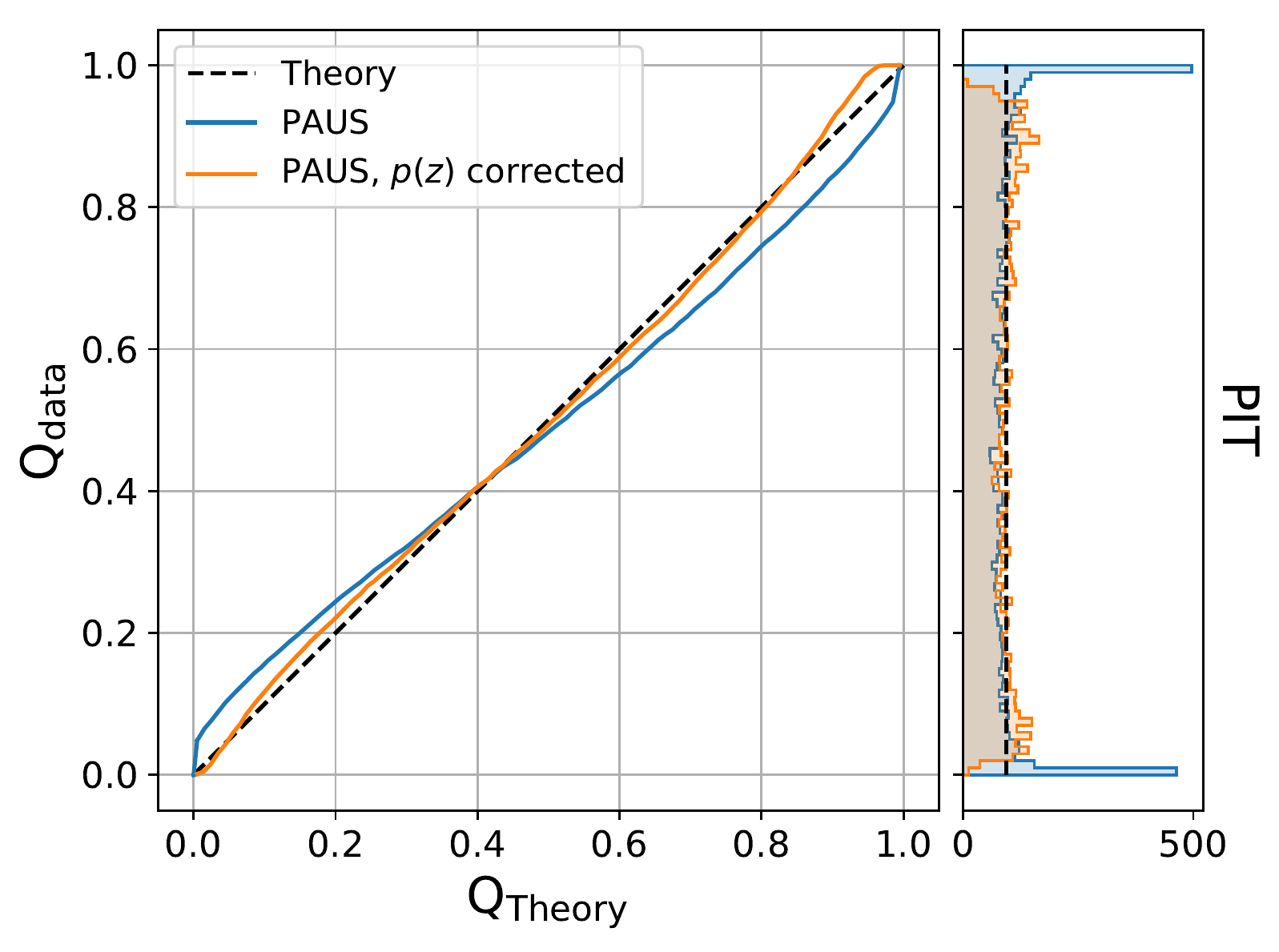}
\end{center}
\caption{The Quantile-Quantile (QQ) plot, which tests the pdfs.  This
is plotted without and with a modified $p(z)$ that accounts for
outliers. The right panel shows the distribution of cumulative pdf
values (PIT), which should be uniform for an accurate pdf.}
\label{qq_plot}
\end{figure}

Several codes, including \textsc{bpz}, \textsc{lephare}, \textsc{annz2}
and \textsc{skynet}, produce pdfs. Depending on the code and data set,
these can either be too broad or narrow \citep{Tanaka2018}. One
approach to quantify the validity of the pdfs is to evaluate the
cumulative of each $p(z)$ at the spectroscopic redshift. By convention
in the photo-z community, we name this the probability integral transform
\citep[PIT][]{Dawid1984}. For a galaxy, this is defined as

\be
\mathrm{PIT} \equiv \int_0^{z_\mathrm{s}} \mathrm{d}z\, p(z),
\ee

\noindent
integrating the pdf from zero to the spectroscopic redshift ($z_\mathrm{s}$).
If the pdfs are correctly estimated, then the PIT of a catalogue will form a
uniform distribution. One way to present the PIT values is the
Quantile-Quantile (QQ) plot. This shows for each quantile (x-axis) of the pdfs
the fraction of the spectroscopic redshifts that is found there. Ideally the
line would fall on the diagonal.

Figure \ref{qq_plot} shows a Quantile-Quantile (QQ) plot for the PAUS
photo-z. The line PAUS use the $p(z)$ directly from the photo-z code (no
corrections). Here the line is lying below and above the diagonal at low and
high quantiles, respectively. The distribution of PIT values (shown in the
right panel) is quite uniform, but the very low and high quantiles have more
galaxies than expected. 

An assumption in the photo-z code is that the data are normally 
distributed (Eq. \ref{chi2_eq}). Unfortunately, the PAUS data reduction
has outliers, e.g. from scattered light and uncorrected cross-talk
\citep{PAUcalib, PAUphoto}. These translate into a different
contribution to the $p(z)$ that is not accounted for in the pdf. The
spikes are caused by photo-z outliers.

A simple model to correct the pdfs is by adding an additional uniformly
distributed contribution, $p_{\mathrm{Outlier}}(z)$, to the distribution

\be
p_{\mathrm{Corrected}}(z) = (1-\kappa)\, p(z) + \kappa\, p_{\mathrm{Outlier}}(z).
\label{kappa_def}
\ee

\noindent
This represents the probability ($\kappa$) that a galaxy is found at a
random location in the redshift fitting range. While there exist more
complex ways of correcting the pdfs \citep{Bordoloi2010}, this model is
sufficient, since we only need to correct for catastrophic outliers.

Note that this correction will also depend on the photo-z quality cut. The
"PAUS, $p(z$) corrected" line in Figure \ref{qq_plot} corresponds to setting $\kappa
= 0.13$ which achieves the smallest differences between the PIT distribution
and the expected values for the 10 and 90 quantiles (peaks in Figure
\ref{qq_plot}). This produces a pdf lying closer to the diagonal and corrects
the PIT values on the edges.

\section{Additional results}
\label{photoz_trends}
\subsection{Redshift dependence}
\begin{figure}
\begin{center}
\xinclude{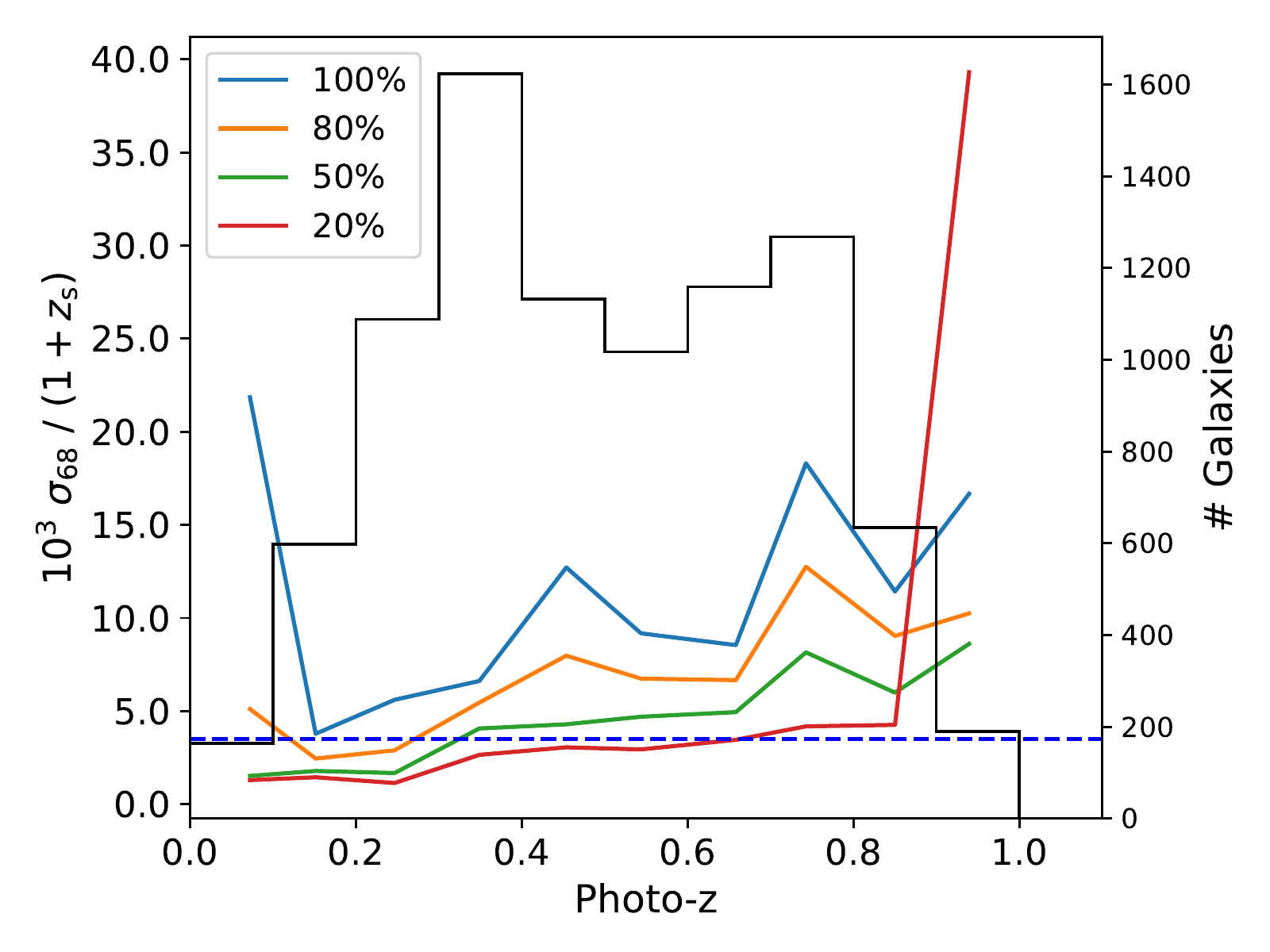}
\end{center}
\caption{The redshift precision as a function of photometric redshift. The $i_\mathrm{AB} <
22.5$ sample is split into 20 bins with equal number of galaxies, before being
split again based on a quality cut (Qz). A horizontal line at $0.0035(1+z)$
shows the nominal PAUS target photo-z precision for a 50\% quality cut. The
black histogram shows the redshift distribution without quality cuts.}
\label{redshift_dependence}
\end{figure}

Figure \ref{redshift_dependence} shows the photo-z scatter when splitting into
redshift bins using the sample with $i_{\mathrm{AB}} < 22.5$. Here the splitting is
based on the photometric redshift, since this is how one will divide a sample
without spectroscopic redshifts. There is a clear increase in the scatter, both
with redshift and fraction of remaining galaxies. At redshift $\sim 0.28$ the
$\text{H}_{\mathrm{\alpha}}$ line disappears from the PAUS wavelength range, leading to a
photo-z degradation. A similar effect happens at $0.69 \lesssim z \lesssim 0.73$,
where OIII and $\text{H}_{\beta}$ leave. A horizontal line indicates
the $0.0035(1+z)$ nominal target for 50\% of the sample. While the photo-z
performance degrades with redshift, the median redshift is low, so the sample
average has a better redshift scatter than the figure might indicate. Furthermore,
at high redshift the 20\% line increases drastically. This is caused by outliers
being scattered to high redshift, but having a narrow $p(z)$, leading to a
good quality parameter.

\subsection{Spatial variations}

\begin{figure}
\begin{center}
\xinclude{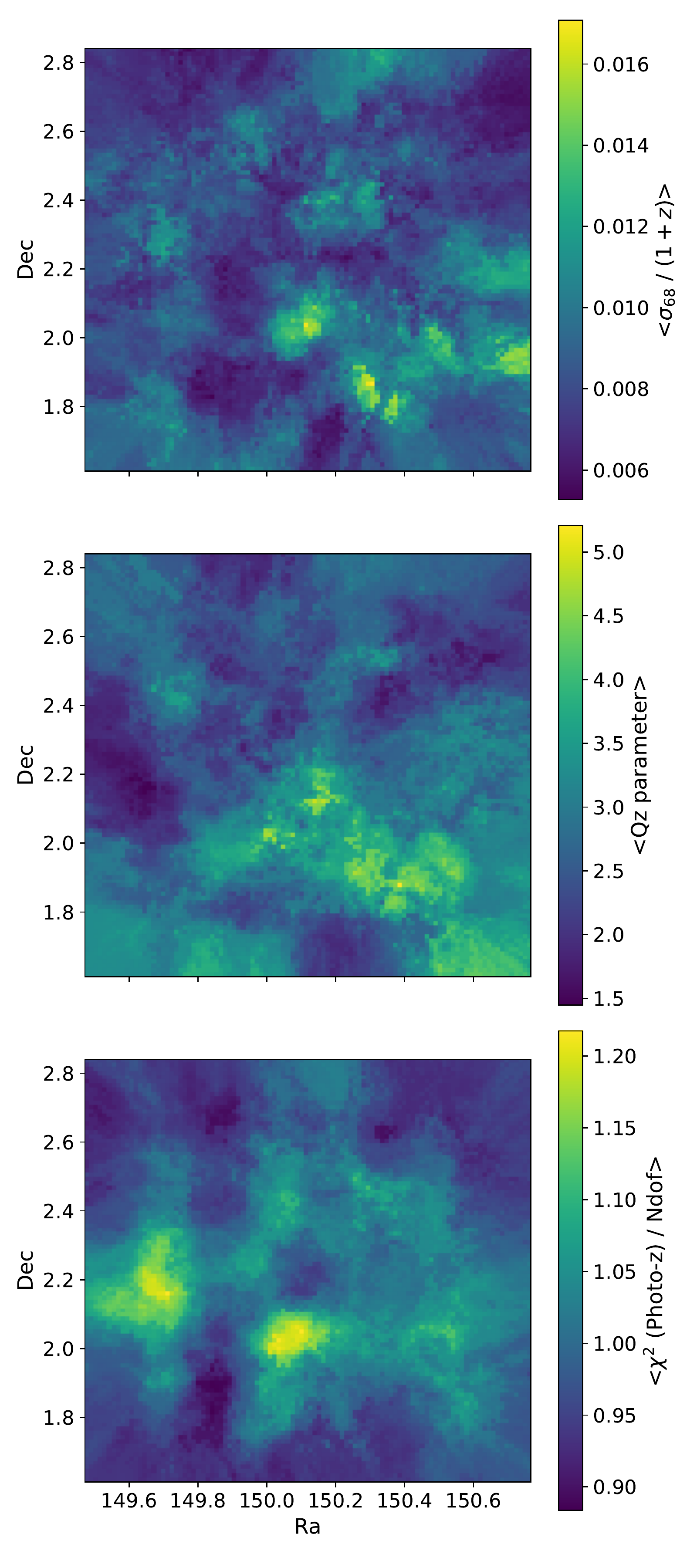}
\end{center}
\caption{The spatial variations of photo-z precision, photo-z quality Qz and
photo-z $\chi^2$ per degree of freedom (Ndof) within the COSMOS field. The
images are generated by associating each pixel with the nearest 200 galaxies.}
\label{spatial_variations}
\end{figure}

Figure \ref{spatial_variations} shows the spatial variations within the COSMOS
field, with each subplot consisting of 100x100 pixels. There are too few
galaxies ($\sim 10000$) in our sample to directly bin these based on position.
Instead, we select the nearest 200 galaxies to each pixel using the tree-based
algorithm in \textsc{scipy}. This roughly correspond to galaxies within 0.09
degrees. Based on this subsample we calculate different forms of statistics
associated to the pixels.

The top panel (Fig. \ref{spatial_variations}) shows the photo-z scatter. Note 
that the value of $\sigma_{68} / (1+z)$ is plotted without any quality cuts. Without
quality cuts the absolute value is higher, but comparable to previous results
for the full sample (Table \ref{prior_table}). Some regions (see colourbar)
have a higher scatter, which can be up to three times higher than in other
regions. This can have implications for the science if not properly accounted
for \citep{Crocce2016}.

In the middle panel the Qz parameter is shown. This form of diagnostics was
previously used in \citep{Marti2014} using ODDS. The Qz parameter is the
default parameter when applying a quality cut (see \S \ref{quality_cuts}) and
smaller values are better. As discussed (\S \ref{quality_cuts}), cutting based
on photo-z quality will introduce inhomogeneities. Last, the bottom panel shows
the $\chi^2$ value when performing the photo-z fit. For each galaxy we use the
minimum for all batches and redshifts. In this plot there is a clear pattern.

\subsection{Emission line strength}

\begin{figure}
\begin{center}
\xinclude{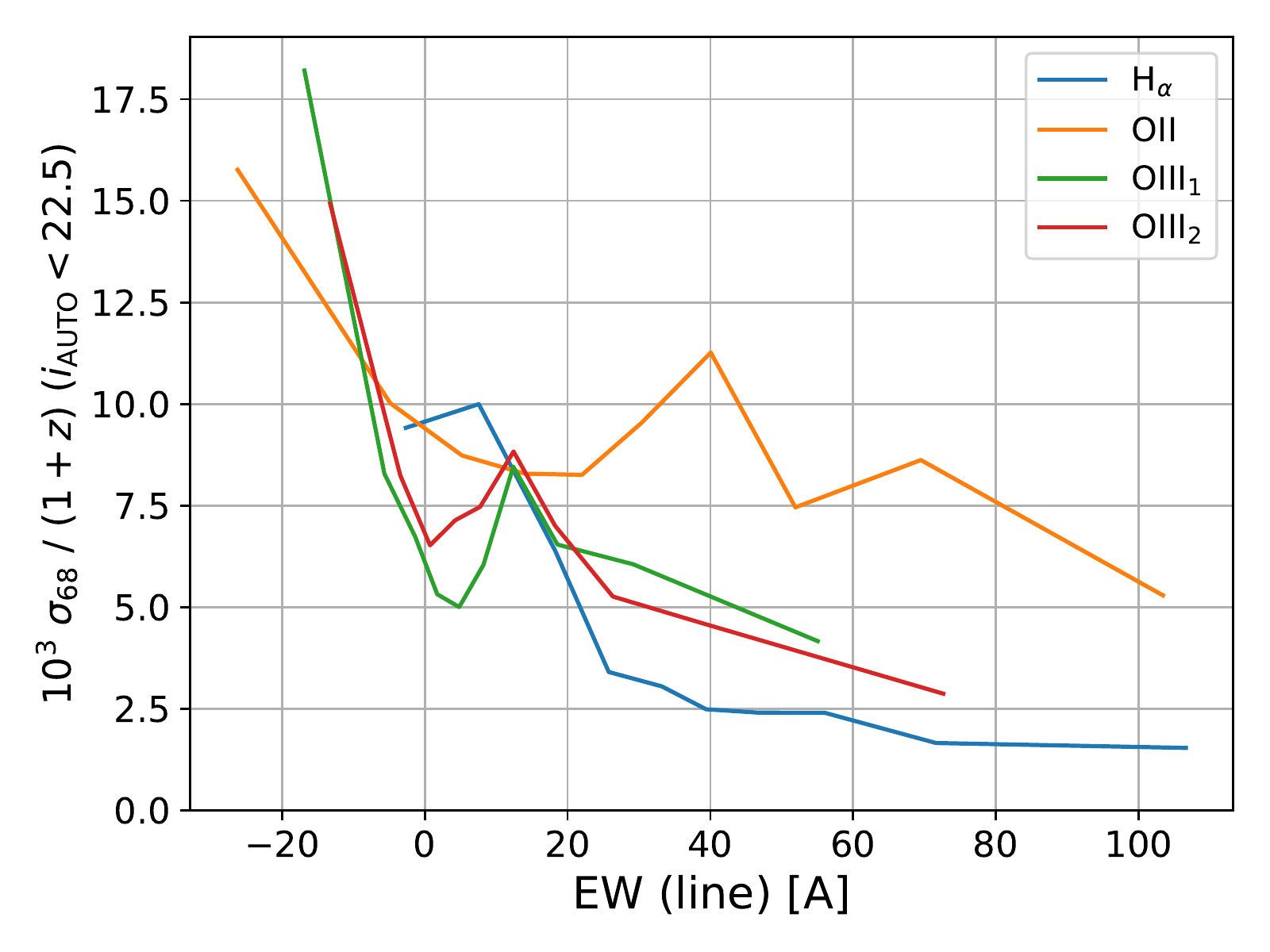}
\end{center}
\caption{The photo-z precision as a function of the equivalent width (EW), for
different emission lines. The x-axis shows the equivalent width for the
narrow band where the emission line has the highest contribution.}
\label{fig:emission_strength}
\end{figure}

Figure \ref{fig:emission_strength} shows $\sigma_{68} / (1+z)$ as a function of
the equivalent width
\be
\mathrm{EW} \equiv 100 \mathrm{\AA}\, (f^\text{Obs} - f^\text{Cont}) / f^{\text{Cont}},
\ee

\noindent
where the narrow-bands are approximated with a 100\AA\ wide top-hat filter and
$f^\text{Obs}$ is the observed flux. The continuum ($f^{\text{Cont}}$)
contribution is estimated by fitting the model used for the photo-z estimation
at the true redshift. For higher emission line strengths, $\sigma_{68} / (1+z)$
decreases for all lines. This shows that emission lines are important for
achieving high photo-z precision with PAUS. A negative emission line strength
occurs when overestimating the continuum, e.g.  by underestimating the
extinction. It can also occur when the estimated flux in the emission line band
is an outlier, e.g. from negative cross-talk. Statistically this yields high
photo-z scatter. For the cases where this happens, we hope to solve this in
future data reductions.

\subsection{Galaxy subsamples}
Luminous red galaxies (LRGs) constitute a useful sample for galaxy clustering
studies. These galaxies are highly clustered, leading to a higher SNR in 2pt
statistics \citep{Eisenstein2001}. They have proven to be an interesting
component in the PAUS galaxy population at $z > 0.4$ \citep{Tortorelli2018}.
Furthermore, their pronounced 4000\AA\ break leads to high photometric
precision \citep{Rozo2016} which makes them a useful sample for many studies,
including BAO, galaxy-galaxy lensing, intrinsic alignments, to name a few
\citep[e.g.][]{Tegmark2006, Joachimi2011, Mandelbaum2013, vanUitert2015,
ElvinPoole2017, Prat2017}.

\begin{figure}
\xinclude{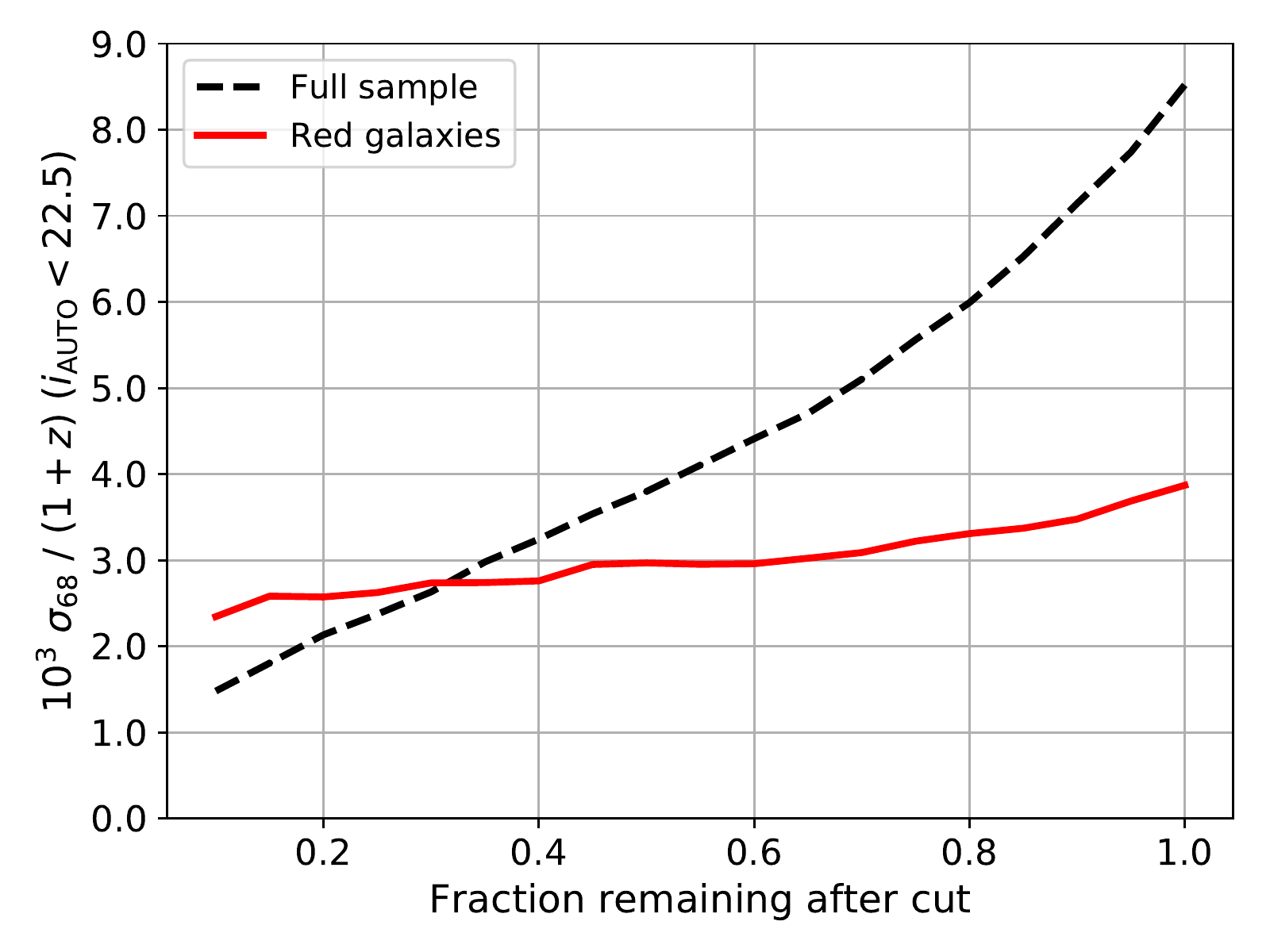} 
\caption{Photo-z precision ($\sigma_{68}/(1+z)$) for different fractions
remaining on a quality cuts. One line shows the precision for the full sample
(dashed), while the other for selected LRGs (solid). The LRGs are selected 
by having a minimal $\chi^2$ for ellipticl templates (run \#1).}
\label{fig:lrg}
\end{figure}

Figure \ref{fig:lrg} shows values of $\sigma_{68}/(1+z)$ for LRGs (solid),
compared to the full sample (dashed). The x-axis shows the remaining fraction
of galaxies selected by cutting on a quality parameter (Qz). The LRGs are
selected by finding galaxies having a minimal $\chi^2$ for run \#1
(Table \ref{sed_combinations}). The LRG sample has a median $i_\mathrm{AUTO}$
of 21.5, which is brighter than the main sample, which has a median
$i_\mathrm{AUTO}$ of 22.1. However, as these are intrinsically bright
galaxies, this sample has a median photometric redshift of 0.69 and extends
out to redshift 1.2. There are not enough spectra above $z>1$ to quantify the
redshift precision, but we expect it to degrade significantly as the 4000\AA\
break is not visible in PAUS beyond $z\sim 1.1$.

\section{Conclusions}
The PAUS survey is an extensive survey currently performed at the William
Herschel Telescope. The novel aspect of the PAUCam instrument is the use of a
40 narrow-band filter set, spaced at 100\AA\ intervals and covering 4500\AA\ to
8500\AA.

The goal is to combine the PAUS narrow bands with deeper broad bands over wide
area weak lensing fields, such as the Canada-France Hawaii Telescope
(CHFT/MegaCam) CFHTLenS Survey \citep{Heymans2012}, Kilo-Degree Survey (KiDS)
\citep{Kuijken2015} or Dark Energy Survey (DES) surveys \citep{DES2005}.

In this paper we focus on COSMOS, which PAUS targeted for science verification,
to quantify the performance of PAUS using actual data. In the case of the
COSMOS field there are many existing measurements with different filters. Of
particular interest are measurements presented in \citet{Laigle2016} with over
32 different broad and intermediate bands, that have been calibrated to measure
the most accurate photo-z values to date.

As a test study, we combine the new PAUS images with only 6 of the COSMOS2015
broad bands representative of the CFHTLenS fields. These are: $u^*$ band data
from the CFHTLenS and \bi{B}, \bi{V}, \bi{g}, \bi{r}, $i^+$ broad bands from
Subaru, obtained in the COSMOS2015 Survey. Thus we have a total of 40+6 filters
in PAUS, while COSMOS2015 used 32, but with wider wavelength coverage. The
COSMOS2015 $i$-band catalogue is used to do forced photometry over the lower
signal-to-noise (SNR) PAUS narrow band images. 

One of the challenges for the PAUS photo-z code is the combination of a few
(six) high SNR bands with many (40) narrow bands with low SNR. Another challenge is
the relative calibration of these surveys, which is validated in \S
\ref{sec:zero_points}. This paper presents the first PAUS photometric redshifts
on the COSMOS field to magnitudes $i$-band $<22.5$. The photometric redshifts
are estimated by a new photo-z code, \textsc{bcnz2}, presented in \S
\ref{photoz_code}. This code is similar to \textsc{eazy} \citep{Brammer2008},
which computes a linear combination of SED templates. However, it has a
different treatment of emission lines and extinction.

Figure \ref{main_photoz_cum} is the main result of this paper. The panels show
the preliminary PAUS photo-z accuracy $\sigma_{68}$ and the outlier
fraction as a function of cumulative $i$-band magnitude. These preliminary result
already match the expected photo-z precision of $\sigma_{68} / (1+z) \simeq
0.0035$ for $i_{\mathrm{AB}} < 22.5$ and a best 50\% photometric redshift
quality cut. The results are also significantly better, for the same objects,
than the state-of-the-art. COSMOS2015 photo-z results are based on
measurements with a much larger wavelength coverage and better signal-to-noise
ratio, but not as good wavelength resolution as PAUS. 

We also find better than expected photo-z accuracy (comparable to spectroscopy) for
high SNR measurements, for emission line galaxies and for colour-selected
subsamples. These results demonstrate the feasibility of the PAUS programme, but
they are neither final nor optimal. When we split the sample in differential
magnitude bins or look at the consistency of the cumulative redshift
probabilities (pdf), we find evidence for an excess of outliers that require
further optimisation and investigation. We are also working on several
improvements to our processing and photo-z codes. We are therefore hopeful
to achieve better performance and present new science applications in the
near future.

\section*{Acknowledgement}
Funding for PAUS has been provided by Durham University (via the ERC StG
DEGAS-259586), ETH Zurich, Leiden University (via ERC StG ADULT-279396 and
Netherlands Organisation for Scientific Research (NWO) Vici grant 639.043.512)
and University College London. The PAUS participants from Spanish institutions
are partially supported by MINECO under grants CSD2007-00060, AYA2015-71825,
ESP2015-66861, FPA2015-68048, SEV-2016-0588, SEV-2016-0597, and MDM-2015-0509,
some of which include ERDF funds from the European Union. IEEC and IFAE are
partially funded by the CERCA program of the Generalitat de Catalunya. The PAUS
data center is hosted by the Port d'Informaci\'{o} Cient\'{i}fica (PIC), maintained
through a collaboration of CIEMAT and IFAE, with additional support from
Universitat Aut\`{o}noma de Barcelona and ERDF. This project has received
funding from the European Union’s Horizon 2020 research and innovation
programme under grant agreement No 776247.

P. Norberg acknowledges the support of the Royal Society through the award of a
University Research Fellowship and the Science and Technology Facilities
Council [ST/P000541/1]. H. Hildebrandt is supported by Emmy Noether (Hi
1495/2-1) and Heisenberg grants (Hi 1495/5-1) of the Deutsche
Forschungsgemeinschaft as well as an ERC Consolidator Grant (No. 770935).

Based on observations obtained with MegaPrime/ MegaCam, a joint project of CFHT
and CEA/IRFU, at the Canada-France-Hawaii Telescope (CFHT) which is operated by
the National Research Council (NRC) of Canada, the Institut National des
Science de l'Univers of the Centre National de la Recherche Scientifique (CNRS)
of France, and the University of Hawaii. This work is based in part on data
products produced at Terapix available at the Canadian Astronomy Data Centre as
part of the Canada-France-Hawaii Telescope Legacy Survey, a collaborative
project of NRC and CNRS.

This work has made use of CosmoHub.
CosmoHub has been developed by the Port d'Informaci\'{o} Cient\'{i}fica (PIC),
maintained through a collaboration of the Institut de F\'{i}sica d'Altes Energies
(IFAE) and the Centro de Investigaciones Energ\'{e}ticas, Medioambientales y
Tecnol\'{o}gicas (CIEMAT), and was partially funded by the "Plan Estatal de
Investigaci\'{o}n Cient\'{i}fica y T\'{e}cnica y de Innovaci\'{o}n" program of the Spanish
government.

\appendix
\section{The \textsc{bcnz} photo-z code}
\label{app_photoz_code}
\subsection{Minimisation algorithm}
The minimisation of the $\chi^2$ (Eq. \ref{main_pz_eq}) has a closed form
solution. However, this includes solutions where some of the amplitudes
(${\bm \alpha}$) are negative. These are undesirable because they lead to unphysical
solutions. Applying a negative amplitude to some SEDs would cancel out features
of the data, leading to worse redshift accuracy. We therefore require the
amplitudes to be positive.

To minimize the $\chi^2$, we used a method for non-negative quadratic
programming, given in \citet{Sha2007}. The minimisation uses an iterative
algorithm, which defines

\begin{align}
A_{xy} \equiv \sum_i \frac{f_i^x f_i^y}{\sigma_i^2}, \quad 
b_{x} \equiv \sum_i \frac{f_i^x \tilde{f}_i}{\sigma_i^2}
\end{align}

\noindent
for templates $x,y$, where the summations are over the bands denoted by $i$.
If ${\bm \alpha}$ is the set of amplitudes at a certain step, the
updated amplitudes $\bar{\bm \alpha}$ at the next step are then

\begin{align}
m_x = \frac{b_x}{\sum_{xy} A_{xy} {\bm \alpha}_y}, \quad
\bar{{\bm \alpha}}_x = m_x {\bm \alpha}_x,
\end{align}

\noindent
where the summation in the determination could use a matrix product. In the
implementation the minimum is estimated at the same time for a set of galaxies,
for all the different redshift bins.

\label{code_details}
\subsection{Language}
The \textsc{bcnz2} code is mainly written in \textsc{python} \citep{Rossum1995},
but with the core algorithm in \textsc{julia} \citep{Bezanson2017}. The
\textsc{python} language is widely used in the astronomical community,
partly because of being a high-level language, allowing to code up difficult
problems in fewer lines. In particular, the \textsc{bcnz2} code relies
heavily on \textsc{pandas} \citep{McKinney2010} and \textsc{xarray}
\citep{Hoyer2017}.

Python code written in the style of \textsc{c} and \textsc{fortran}, relying on
loops, is slow. For numerical tasks, one should either use fast building blocks
as matrix operations or call a library written in another language.
Alternatively one can use \textsc{numba} \citep{Lam2015}, a just-in-time
compiler converting math intensive Python to machine instructions. Adding a
single line \textsc{numba} decorator (@numba.jit) reduced the runtime to about
2/3 of the original value. Other alternatives include \textsc{cython}
\citep{Behnel2010}, \textsc{c++} \citep{Stroustrup2000} or \textsc{julia}. In
the end we decided on \textsc{julia}, since the code was readable and executed
fast.

\subsection{Infrastructure}
Running the photo-z code can be time consuming. Having access to an environment
with multiple CPUs allows us to calculate the photo-z faster, allowing for more
iterations. The \textsc{bcnz2} code is integrated within the Apache Spark
cluster \citep{Zaharia2010} running at Port d'Informaci\'{o} Cient\'{i}fica
(PIC). This platform is also used for CosmoHub \citep{Carretero2017}. Spark is
suitable for programs where the calculations can be split into independent
parts, with the result being combined at the end (map-reduce). For the photo-z
we split into sets of galaxies. Users can either run \textsc{bcnz2} locally or
remotely run the code at PIC.

\subsection{Convergence}
\label{code_convergence}

\begin{figure}
\begin{center}
\xinclude{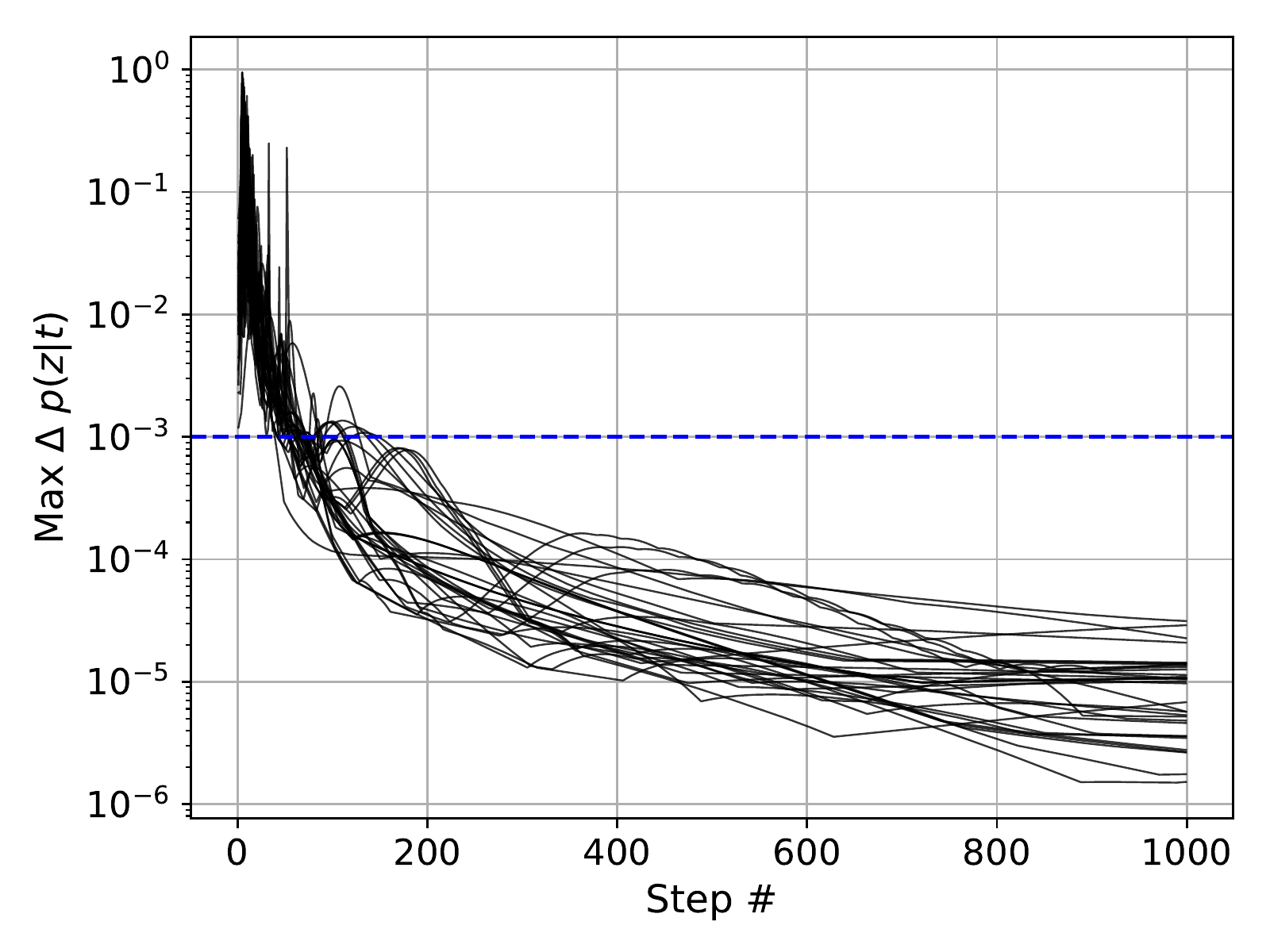}
\end{center}
\caption{A convergence test, showing the maximum absolute change in p(z) for
all redshifts for a set for 10 galaxies. On the x-axis is the number of steps
in the iterative minimisation, while each line corresponds to a photo-z run
(Table \ref{sed_combinations}).}
\label{fig_convergence}
\end{figure}

The basic minimisation algorithm is made to minimise the $\chi^2$ separately at
each redshift, and is proven to reach convergence \citep{Sha2007}, but the
question is how fast convergence is reached. That is important when running the
photo-z, since the minimisation is the most time consuming part.

Figure \ref{fig_convergence} shows a benchmark for the convergence. On the
x-axis is the iteration step, while the y-axis shows the maximum absolute
change in $p(z)$. The maximum absolute change in $p(z)$ is estimated between
two iteraction and selects the redshift with maximum change for any of the
galaxies in the batch of 10 galaxies.

This quantity was chosen, since it relates more directly to
the error we want to minimise. We first attempted to study the convergence
looking at the model amplitude (${\bm \alpha}$) changes. This had the problem of some
amplitudes being unconstrained, e.g. when an emission line does not enter into
any of the bands. Further, focusing on the $\chi^2$ value is also problematic,
since changes to high $\chi^2$ values are less important for the final result.
Hence we ended up focusing on the $p(z)$ values.

In this plot each line corresponds to one of the 45 photo-z runs (Table
\ref{sed_combinations}). Here we selected 10 galaxies, which correspond to how
many are usually being run together. For each step we estimated the $p(z)$ for
that photo-z run. Note that most runs do not correspond to the optimal.  The
distribution will therefore be broader and the convergence slightly slower.

While the $\chi^2$ is proven to converge uniformly, this is not the case for
the $p(z)$. During the minimisation, the $\chi^2$ at different redshift grid
values will converge faster to the correct value. In practice, we find many
cases where the $p(z)$ peak position changes from one redshift to another after
some iterations. This explains why some of the lines increase within the first
iterations ($<200$). A horizontal line marks a very stringent requirement on
the convergence. By default we run all batches with 1000 iterations, although
500 should be sufficient.

\section{Miscellaneous}
\label{add_results}
\paragraph*{Spectroscopic completeness:}
\begin{figure}
\xinclude{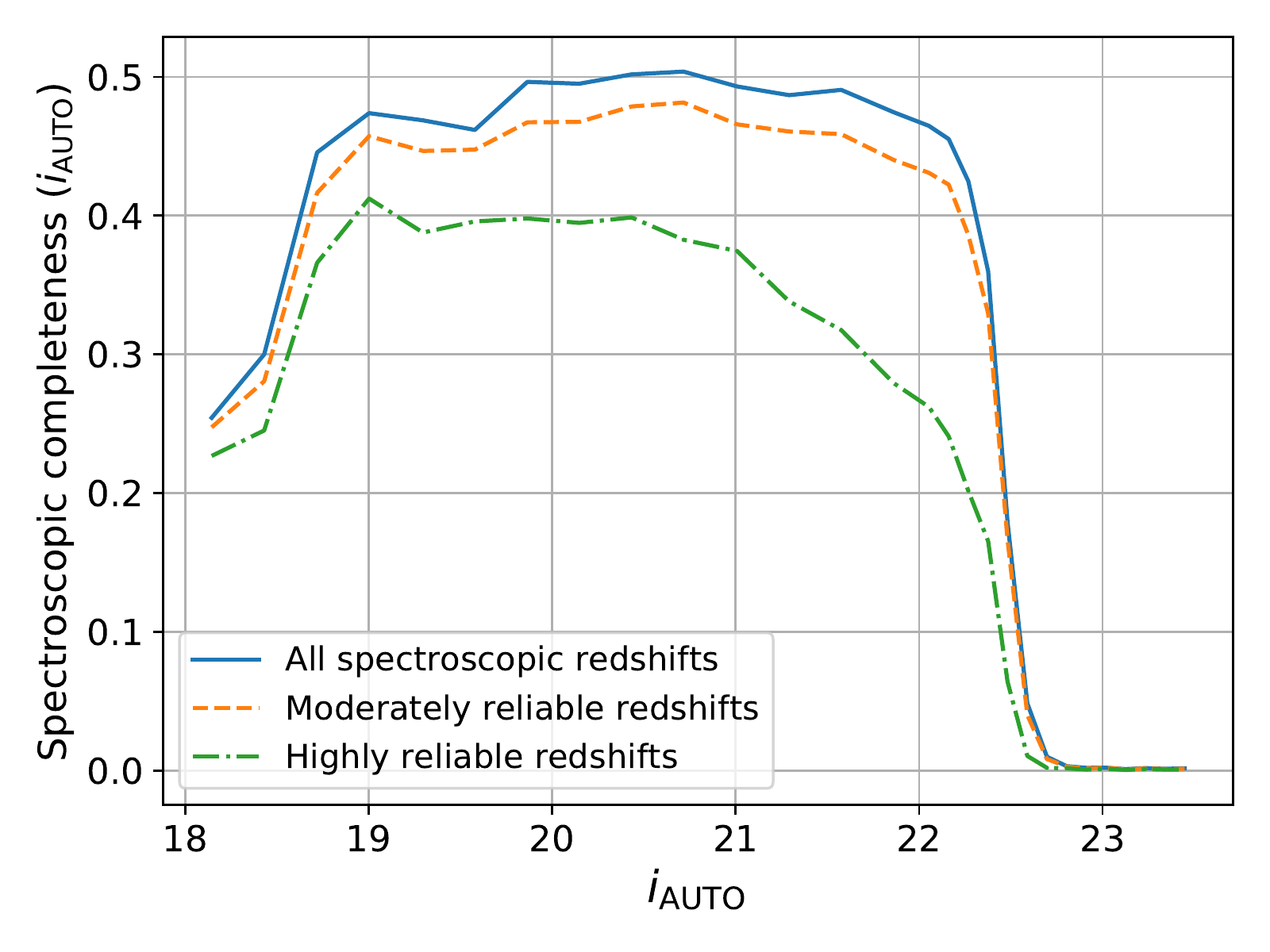}
\caption{The completeness in the zCOSMOS DR3 bright sample. Here the
completeness is the fraction of galaxies with spec-z compared to the full
COSMOS sample for different magnitude bins. Three lines show the full sample
(solid), when selecting moderately secure redshifts (dashed) [classes: 3.x,
4.x, 2.5, 2.4, 1.5, 9.5, 9.4, 9.3, 18.5, 18.3] and highly secure redshifts
(dash-dotted) [classes: 3.x, 4.x].}
\label{fig:cosmos_completeness}
\end{figure}

Figure \ref{fig:cosmos_completeness} shows the spectroscopic redshift
completeness as a function of $i_\mathrm{AUTO}$, the SExtractor's AUTO
magnitude (MAG\_AUTO) in the $i$-band. zCOSMOS DR3 bright data have a 44\%
completeness for $i_\mathrm{{AUTO}}\leq22.5$, which reduces to 28\% after
imposing the spectra to be highly reliable. For this paper we use the highly
reliable redshifts (3.x, 4.x), as suggested in zCOSMOS DR3\footnote{zCOSMOS
DR3 release note.}. The spectroscopic completeness of our reference has to be
kept in mind when presenting the result as a function of magnitude.

\paragraph*{Broad band transmission curves:}
\begin{figure}
\xinclude{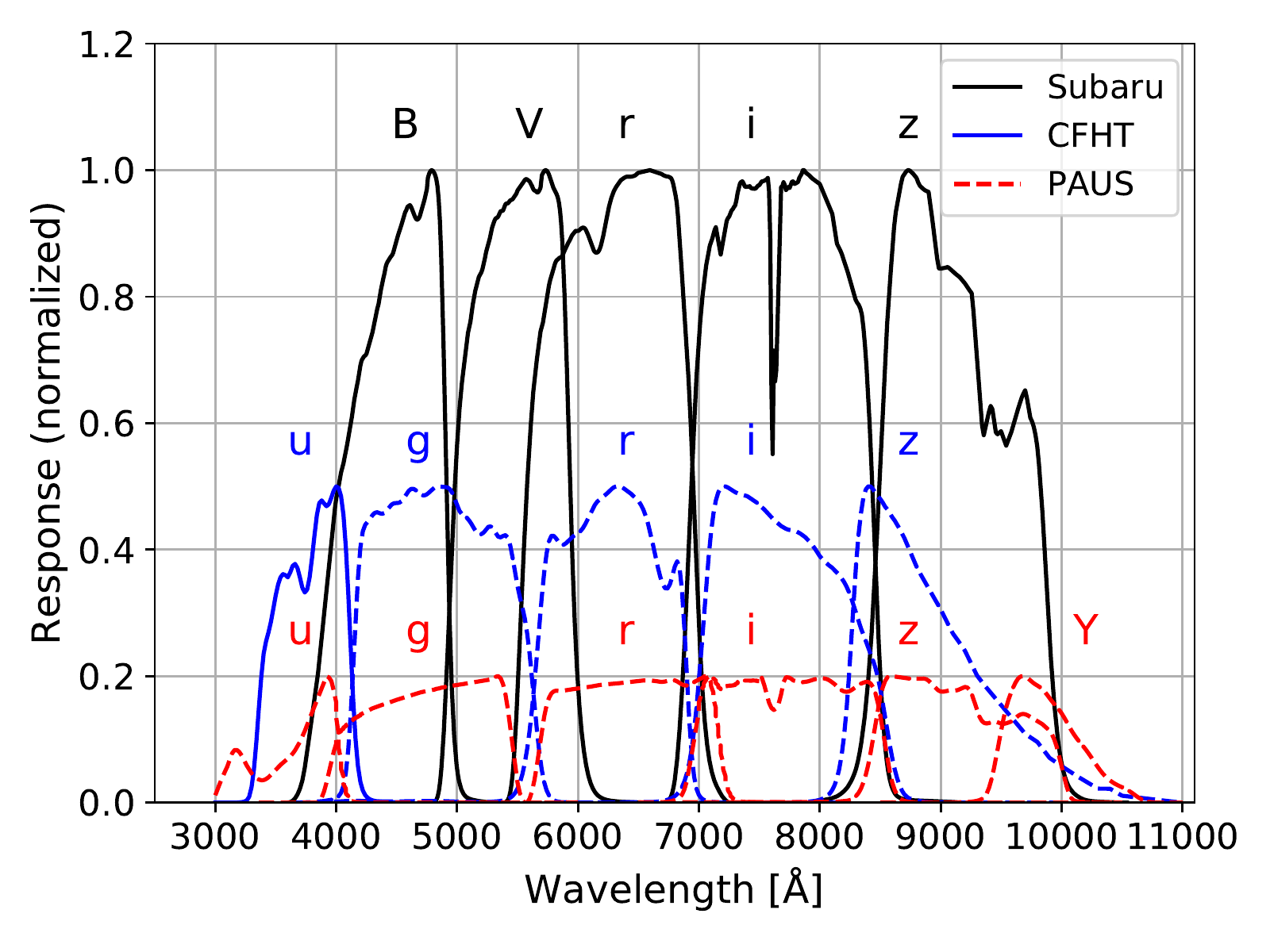}
\caption{The Subaru, CFHT and PAUS broad band throughputs, which for improving
the visualisation are each normalised to 1.0, 0.5 and 0.2, respectively. Solid
lines mark filters which are used in the photo-z run, while dashed filters are
only included as a reference.}
\label{bb_filters}
\end{figure}

Figure \ref{bb_filters} shows the broad band transmission curves. PAUS science
cases use external broad band datasets as reference catalogues, for which we
produce precise photo-zs. Therefore PAUCam's own broad bands are not used
currently in PAUS, but are included for reference. The transmission for broad
bands corresponding to Subaru and the Canadian-France Hawaii Telescope
camera\footnote{The Subaru and CFHT filter curves were downloaded from
http://cosmos.astro.caltech.edu/page/filterset} are shown.

\paragraph*{Photo-z scatter}
\begin{figure}
\begin{center}
\xinclude{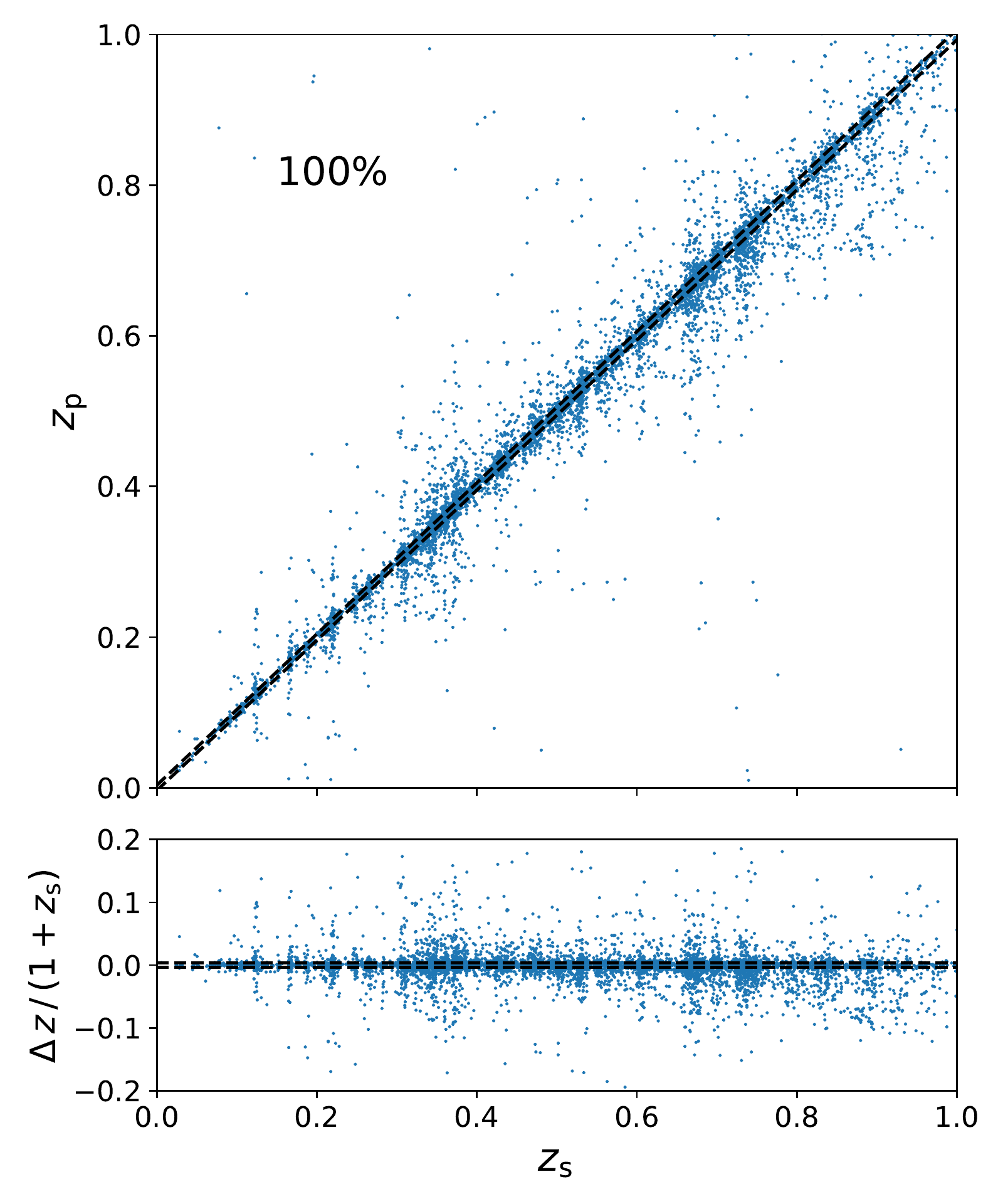}
\end{center}
\caption{The photo-z performance without quality cuts as a function of spectroscopic redshift
($z_\mathrm{s}$). The y-axis top panel shows the photometric redshift
($z_\mathrm{p}$), while the bottom panel shows the scatter $(z_\mathrm{p} -
z_\mathrm{s})  / (1+z_\mathrm{s})$. The two lines mark the $1-\sigma$ limits
for the 0.0035(1+z) target precision.}
 \label{photoz_scatter}
\end{figure}

\begin{figure}
\begin{center}
\xinclude{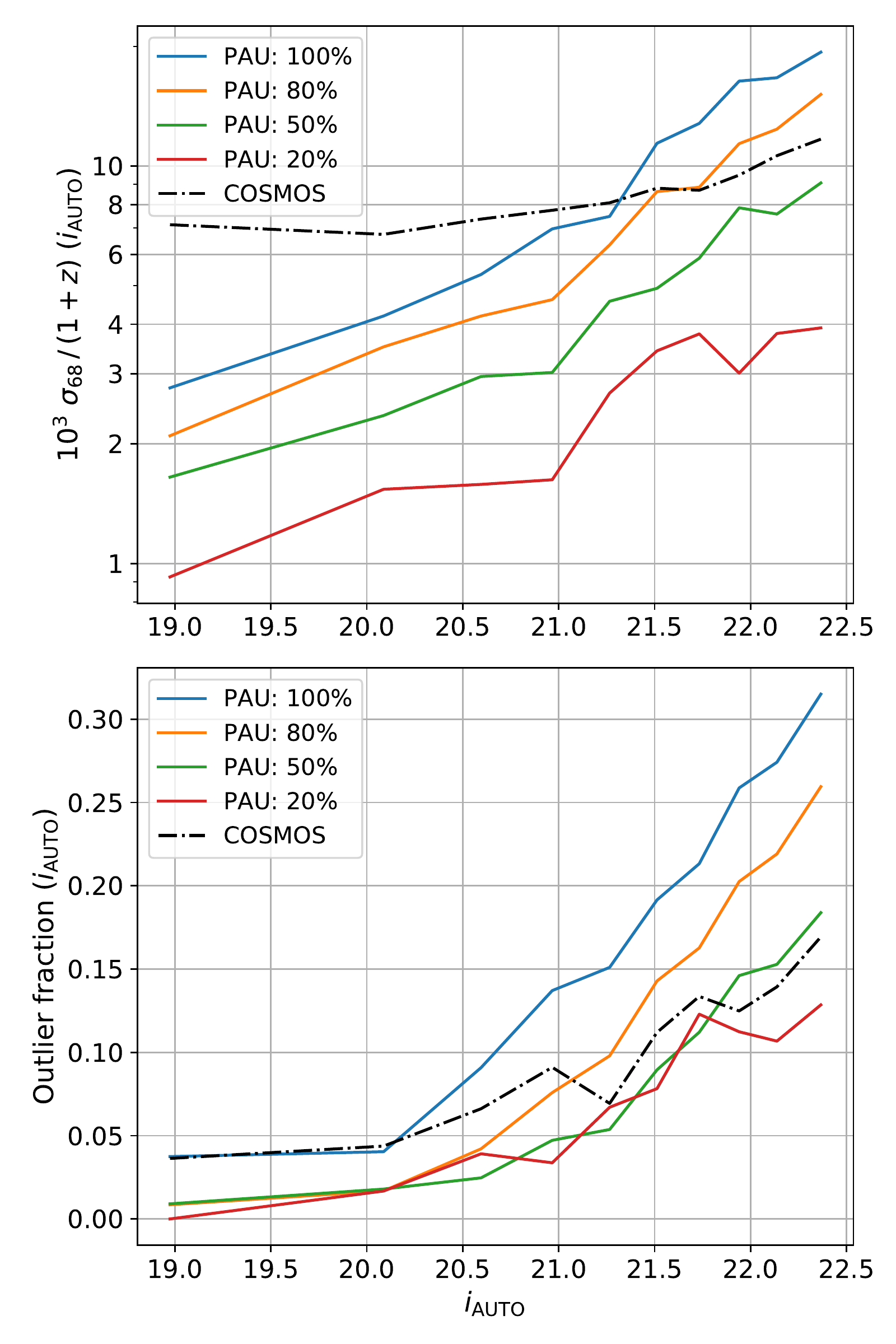}
\end{center}
\caption{Same as Figure \ref{main_photoz_cum}, but showing the differential
performance as a function of magnitude.}
\label{main_photoz_binned}
\end{figure}

Figure \ref{photoz_scatter} shows the photo-z scatter plot, corresponding to
the data in Figure \ref{main_photoz_cum} and \ref{main_photoz_binned}. Here
the result is shown for the full sample (no quality cuts).

\paragraph*{Differential magnitude bins:}
Figure \ref{main_photoz_binned} shows the same results as Figure
\ref{main_photoz_cum}, but in differential, instead of cumulative magnitude
bins. Here we can see more clearly the degradation of both accuracy and number
of outliers at the faint end of the sample. There is an excess of outliers
which arise from both the lower signal-to-noise ratio and the non-optimal
treatment of data and photo-z code in this regime. These are preliminary
results and we are implementing new methods to improve this performance.

\section{Galaxy internal extinction}
\label{dust_extinction}
There are multiple contributions to the extinction, both from the Milky Way and
internally in the galaxies we observe. The Galactic extinction is
corrected for by the calibration procedure described in \citet{PAUcalib}.
Internal galaxy extinction varies from galaxy to galaxy and needs
to be included in the modelling of each galaxy SED.

Galaxy internal extinction results from dust scattering light, reducing the
light transmitted in the direction of the observer. Let $F_i(\lambda)$ be the
intrinsic galaxy spectrum and $F_o(\lambda)$ be the observable spectrum after
extinction. Then

\be
F_o(\lambda) = F_i(\lambda) {10}^{-0.4 E(B-V) k(\lambda)}
\ee

\noindent
relates the two. Here the $E(B-V)$ parameter measures the magnitude
difference in the \bi{B} and \bi{V} bands. As follows, the $k(\lambda)$
wavelength dependent function is determined through observations.

\begin{figure}
\xinclude{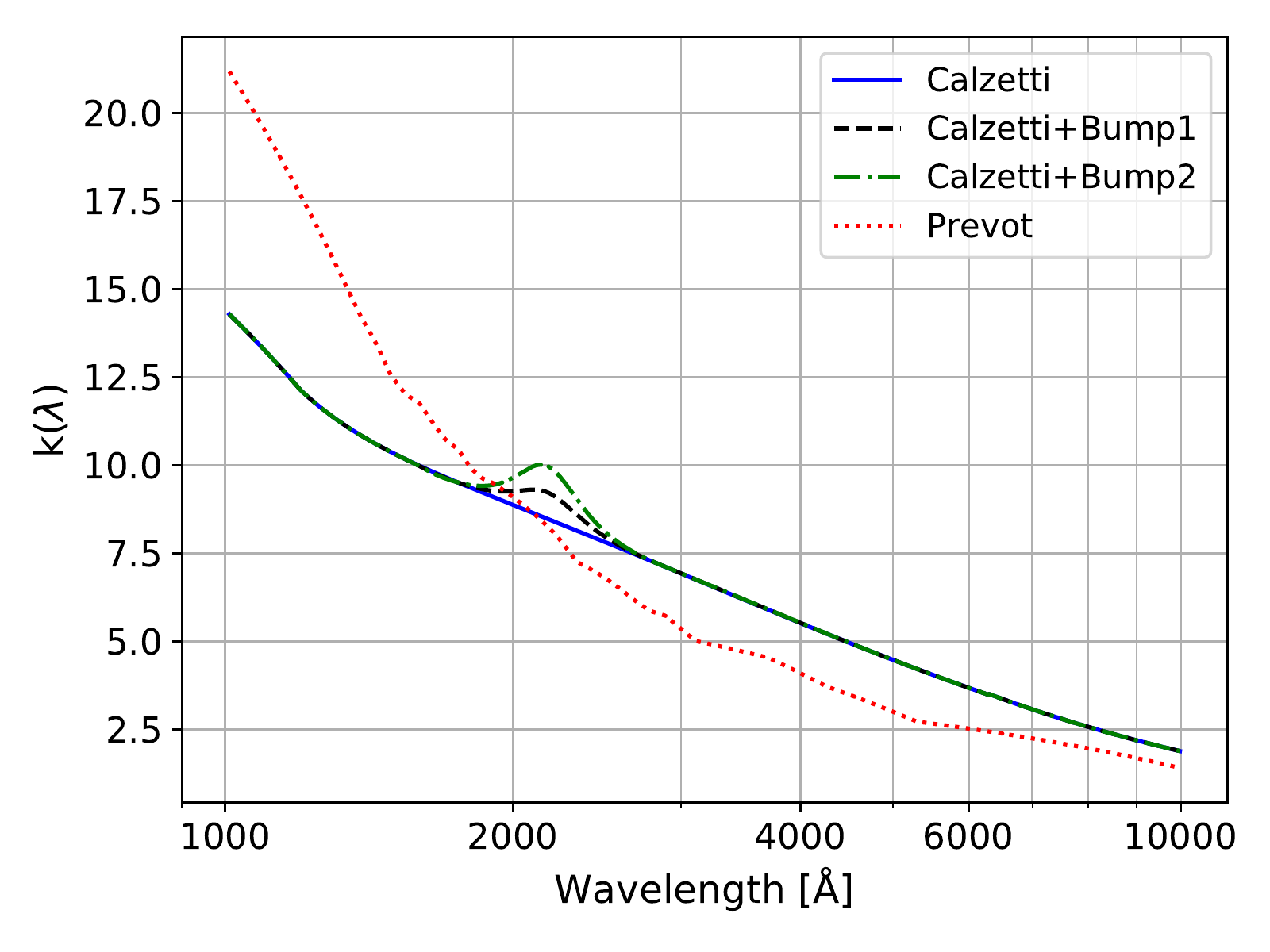}
\caption{The internal extinction curves. The Calzetti extinction curve (solid)
is applied to starburst galaxies. Two extinction laws (dashed and dot-dashed)
add an additional contribution at 2175\AA. The Prevot extinction (dotted) is
applied to spiral galaxies.}
\label{extinction_curves}
\end{figure}

Figure \ref{extinction_curves} shows the wavelength dependence of the
extinction laws used. Multiple relations exist in the literature. One commonly
used is the extinction law of \citet{Calzetti2000}, which is fitted to observed
starburst galaxies. The photo-z code uses the \citet{Calzetti2000} extinction
law for starburst templates.

A characteristic feature in the extinction is the 2175\AA\ bump
\citep{Stecher1965}. Polycyclic aromatic hydrocarbon (PAH) molecular
transitions have been suggested as an explanation, but the origin
is still debated \citep{Xiang2011}. This feature is left out of
the Calzetti relation, since their starburst galaxy spectra did not show
a prominent feature around 2175\AA\ \citep{Calzetti1994}.

In \citet{Fitzpatrick1986, Fitzpatrick2007} the authors parameterised the 
2175\AA\ bump with a Drude profile

\be
D(x, x_0, \gamma) = \frac{x^2}{{(x^2 - x_0^2)}^2 + x^2 \gamma^2},
\ee

\noindent
which gives an analytical expression for the 2175\AA\ bump. The COSMOS2015
paper used the Calzetti extinction law, including this additional contribution.
This paper uses their tabulated values. Two different versions exist, one with
double strengths of the 2175\AA\ feature. When running with the Calzetti law,
we also fit with these two modified versions. The Prevot extinction was
measured in the Small Magellanic Cloud (SMC) \citep{Prevot1984}. This
extinction law is commonly applied for spiral galaxies, including in
COSMOS2015. By default we do not include the Prevot extinction, but have tested
applying it to spiral templates. This significanly reduced the photo-z
precision, hence we do not include the Prevot extinction.

\bibliography{bibpz}{}

\begin{thebibliography}{100}
\expandafter\ifx\csname natexlab\endcsname\relax\def\natexlab#1{#1}\fi

\bibitem[{{Alarcon}, {Eriksen} \& {Gaztanaga}(2018){Alarcon}, {Eriksen}, \&
  {Gaztanaga}}]{Alarcon2018}
{Alarcon} A., {Eriksen} M., {Gaztanaga} E., 2018, \mnras, 473, 1444

\bibitem[{{Allen}(1976)}]{Allen1976}
{Allen} C.~W., 1976, {Astrophysical Quantities}

\bibitem[{{Arnouts} \& {Ilbert}(2011)}]{Arnouts2011}
{Arnouts} S., {Ilbert} O., 2011, {LePHARE: Photometric Analysis for Redshift
  Estimate}. Astrophysics Source Code Library

\bibitem[{{Asorey} {et~al}\mbox{.}(2016){Asorey}, {Carrasco Kind},
  {Sevilla-Noarbe}, {Brunner}, \& {Thaler}}]{Asorey2016}
{Asorey} J., {Carrasco Kind} M., {Sevilla-Noarbe} I., {Brunner} R.~J., {Thaler}
  J., 2016, \mnras, 459, 1293

\bibitem[{{Baldwin}, {Phillips} \& {Terlevich}(1981){Baldwin}, {Phillips}, \&
  {Terlevich}}]{Baldwin1981}
{Baldwin} J.~A., {Phillips} M.~M., {Terlevich} R., 1981, \pasp, 93, 5

\bibitem[{{Beck} {et~al}\mbox{.}(2016){Beck}, {Dobos}, {Yip}, {Szalay}, \&
  {Csabai}}]{Beck2016}
{Beck} R., {Dobos} L., {Yip} C.-W., {Szalay} A.~S., {Csabai} I., 2016, \mnras,
  457, 362

\bibitem[{Behnel {et~al}\mbox{.}(2011)Behnel, Bradshaw, Citro, Dalcin,
  Seljebotn, \& Smith}]{Behnel2010}
Behnel S., Bradshaw R., Citro C., Dalcin L., Seljebotn D., Smith K., 2011,
  Computing in Science Engineering, 13, 31

\bibitem[{{Ben{\'{\i}}tez}(2000)}]{Benitez2000}
{Ben{\'{\i}}tez} N., 2000, \apj, 536, 571

\bibitem[{{Ben{\'{\i}}tez} {et~al}\mbox{.}(2009){Ben{\'{\i}}tez},
  {Gazta{\~n}aga}, {Miquel}, {Castander}, {Moles}, {Crocce},
  {Fern{\'a}ndez-Soto}, {Fosalba}, {Ballesteros}, {Campa}, {Cardiel-Sas},
  {Castilla}, {Crist{\'o}bal-Hornillos}, {Delfino}, {Fern{\'a}ndez},
  {Fern{\'a}ndez-Sopuerta}, {Garc{\'{\i}}a-Bellido}, {Lobo}, {Mart{\'{\i}}nez},
  {Ortiz}, {Pacheco}, {Paredes}, {Pons-Border{\'{\i}}a}, {S{\'a}nchez},
  {S{\'a}nchez}, {Varela}, \& {de Vicente}}]{Benitez2009}
{Ben{\'{\i}}tez} N. {et~al.}, 2009, \apj, 691, 241

\bibitem[{{Bertin}(2011)}]{Bertin2011}
{Bertin} E., 2011, in Astronomical Society of the Pacific Conference Series,
  Vol. 442, Astronomical Data Analysis Software and Systems XX, {Evans} I.~N.,
  {Accomazzi} A., {Mink} D.~J., {Rots} A.~H., eds., p. 435

\bibitem[{{Bertin} \& {Arnouts}(1996)}]{Bertin1996}
{Bertin} E., {Arnouts} S., 1996, \aaps, 117, 393

\bibitem[{Bezanson {et~al}\mbox{.}(2017)Bezanson, Edelman, Karpinski, \&
  Shah}]{Bezanson2017}
Bezanson J., Edelman A., Karpinski S., Shah V.~B., 2017, SIAM Review, 59, 65

\bibitem[{{Bonnett}(2015)}]{Bonnett2015}
{Bonnett} C., 2015, \mnras, 449, 1043

\bibitem[{{Bordoloi}, {Lilly} \& {Amara}(2010){Bordoloi}, {Lilly}, \&
  {Amara}}]{Bordoloi2010}
{Bordoloi} R., {Lilly} S.~J., {Amara} A., 2010, \mnras, 406, 881

\bibitem[{{Brammer}, {van Dokkum} \& {Coppi}(2008){Brammer}, {van Dokkum}, \&
  {Coppi}}]{Brammer2008}
{Brammer} G.~B., {van Dokkum} P.~G., {Coppi} P., 2008, \apj, 686, 1503

\bibitem[{{Bruzual} \& {Charlot}(2003)}]{Bruzal2003}
{Bruzual} G., {Charlot} S., 2003, \mnras, 344, 1000

\bibitem[{{Cabayol} {et~al}\mbox{.}(2018){Cabayol}, {Sevilla-Noarbe},
  {Fern{\'a}ndez}, {Carretero}, {Eriksen}, {Serrano}, {Alarc{\'o}n}, {Amara},
  {Casas}, {Castander}, {de Vicente}, {Folger}, {Garc{\'{\i}}a-Bellido},
  {Gaztanaga}, {Hoekstra}, {Miquel}, {Padilla}, {S{\'a}nchez}, {Stothert},
  {Tallada}, \& {Tortorelli}}]{Cabayol2018}
{Cabayol} L. {et~al.}, 2018, preprint(ArXiv:1806.08545)

\bibitem[{{Calzetti} {et~al}\mbox{.}(2000){Calzetti}, {Armus}, {Bohlin},
  {Kinney}, {Koornneef}, \& {Storchi-Bergmann}}]{Calzetti2000}
{Calzetti} D., {Armus} L., {Bohlin} R.~C., {Kinney} A.~L., {Koornneef} J.,
  {Storchi-Bergmann} T., 2000, \apj, 533, 682

\bibitem[{{Calzetti}, {Kinney} \& {Storchi-Bergmann}(1994){Calzetti}, {Kinney},
  \& {Storchi-Bergmann}}]{Calzetti1994}
{Calzetti} D., {Kinney} A.~L., {Storchi-Bergmann} T., 1994, \apj, 429, 582

\bibitem[{Carretero {et~al}\mbox{.}(2017)Carretero {et~al.}}]{Carretero2017}
Carretero J., {et~al.}, 2017, PoS, EPS-HEP2017, 488

\bibitem[{{Casas} {et~al}\mbox{.}(2016){Casas}, {Cardiel-Sas}, {Castander},
  {Díaz}, {Gaweda}, {Jiménez Rojas}, {Jiménez}, {Lamensans}, {Padilla},
  {Rodriguez}, {Sanchez}, \& {Sevilla Noarbe}}]{Casas2016}
{Casas} R. {et~al.}, 2016, in Proc.SPIE, Vol. 9908, pp. 9908 -- 9908 -- 12

\bibitem[{{Casas} {et~al}\mbox{.}(2014){Casas}, {Cardiel-Sas}, {Castander},
  {Jiménez}, \& {de Vicente}}]{Casas2014}
{Casas} R., {Cardiel-Sas} L., {Castander} F., {Jiménez} J., {de Vicente} J.,
  2014, in Proc.SPIE, Vol. 9147, pp. 9147 -- 9147 -- 8

\bibitem[{{Castander} {et~al}\mbox{.}({in prep.}){Castander}, {Eriksen},
  {Serrano}, \& {et al.}}]{PAUcalib}
{Castander} F., {Eriksen} M., {Serrano} S., {et al.}, {in prep.}

\bibitem[{{Castander} \& et~al.(2012)}]{Castander2012}
{Castander} F., et~al., 2012, in Proc.SPIE, Vol. 8446, pp. 8446 -- 8446 -- 9

\bibitem[{{Catelan}, {Kamionkowski} \& {Blandford}(2001){Catelan},
  {Kamionkowski}, \& {Blandford}}]{Catelan2000}
{Catelan} P., {Kamionkowski} M., {Blandford} R.~D., 2001, \mnras, 320, L7

\bibitem[{{Crocce} {et~al}\mbox{.}(2016){Crocce}, {Carretero}, {Bauer}, {Ross},
  {Sevilla-Noarbe}, {Giannantonio}, {Sobreira}, {Sanchez}, {Gaztanaga},
  {Carrasco Kind}, {S{\'a}nchez}, {Bonnett}, {Benoit-L{\'e}vy}, {Brunner},
  {Carnero Rosell}, {Cawthon}, {Fosalba}, {Hartley}, {Kim}, {Leistedt},
  {Miquel}, {Peiris}, {Percival}, {Rosenfeld}, {Rykoff}, {S{\'a}nchez},
  {Abbott}, {Abdalla}, {Allam}, {Banerji}, {Bernstein}, {Bertin}, {Brooks},
  {Buckley-Geer}, {Burke}, {Capozzi}, {Castander}, {Cunha}, {D'Andrea}, {da
  Costa}, {Desai}, {Diehl}, {Eifler}, {Evrard}, {Fausti Neto}, {Fernandez},
  {Finley}, {Flaugher}, {Frieman}, {Gerdes}, {Gruen}, {Gruendl}, {Gutierrez},
  {Honscheid}, {James}, {Kuehn}, {Kuropatkin}, {Lahav}, {Li}, {Lima}, {Maia},
  {March}, {Marshall}, {Martini}, {Melchior}, {Miller}, {Neilsen}, {Nichol},
  {Nord}, {Ogando}, {Plazas}, {Romer}, {Sako}, {Santiago}, {Schubnell},
  {Smith}, {Soares-Santos}, {Suchyta}, {Swanson}, {Tarle}, {Thaler}, {Thomas},
  {Vikram}, {Walker}, {Wechsler}, {Weller}, {Zuntz}, \& {DES
  Collaboration}}]{Crocce2016}
{Crocce} M. {et~al.}, 2016, \mnras, 455, 4301

\bibitem[{{Croft} \& {Metzler}(2000)}]{Croft2000}
{Croft} R.~A.~C., {Metzler} C.~A., 2000, \apj, 545, 561

\bibitem[{{Davis} {et~al}\mbox{.}(2003){Davis}, {Faber}, {Newman}, {Phillips},
  {Ellis}, {Steidel}, {Conselice}, {Coil}, {Finkbeiner}, {Koo}, {Guhathakurta},
  {Weiner}, {Schiavon}, {Willmer}, {Kaiser}, {Luppino}, {Wirth}, {Connolly},
  {Eisenhardt}, {Cooper}, \& {Gerke}}]{Davis2003}
{Davis} M. {et~al.}, 2003, in \procspie, Vol. 4834, Discoveries and Research
  Prospects from 6- to 10-Meter-Class Telescopes II, {Guhathakurta} P., ed.,
  pp. 161--172

\bibitem[{Dawid(1984)}]{Dawid1984}
Dawid A.~P., 1984, Journal of the Royal Statistical Society. Series A
  (General), 147, 278

\bibitem[{{De Vicente}, {S{\'a}nchez} \& {Sevilla-Noarbe}(2016){De Vicente},
  {S{\'a}nchez}, \& {Sevilla-Noarbe}}]{Vicente2016}
{De Vicente} J., {S{\'a}nchez} E., {Sevilla-Noarbe} I., 2016, \mnras, 459, 3078

\bibitem[{{Driver} {et~al}\mbox{.}(2009){Driver}, {Norberg}, {Baldry},
  {Bamford}, {Hopkins}, {Liske}, {Loveday}, {Peacock}, {Hill}, {Kelvin},
  {Robotham}, {Cross}, {Parkinson}, {Prescott}, {Conselice}, {Dunne}, {Brough},
  {Jones}, {Sharp}, {van Kampen}, {Oliver}, {Roseboom}, {Bland-Hawthorn},
  {Croom}, {Ellis}, {Cameron}, {Cole}, {Frenk}, {Couch}, {Graham}, {Proctor},
  {De Propris}, {Doyle}, {Edmondson}, {Nichol}, {Thomas}, {Eales}, {Jarvis},
  {Kuijken}, {Lahav}, {Madore}, {Seibert}, {Meyer}, {Staveley-Smith},
  {Phillipps}, {Popescu}, {Sansom}, {Sutherland}, {Tuffs}, \&
  {Warren}}]{Driver2009}
{Driver} S.~P. {et~al.}, 2009, Astronomy and Geophysics, 50, 5.12

\bibitem[{{Eisenstein} {et~al}\mbox{.}(2001){Eisenstein}, {Annis}, {Gunn},
  {Szalay}, {Connolly}, {Nichol}, {Bahcall}, {Bernardi}, {Burles}, {Castander},
  {Fukugita}, {Hogg}, {Ivezi{\'c}}, {Knapp}, {Lupton}, {Narayanan}, {Postman},
  {Reichart}, {Richmond}, {Schneider}, {Schlegel}, {Strauss}, {SubbaRao},
  {Tucker}, {Vanden Berk}, {Vogeley}, {Weinberg}, \& {Yanny}}]{Eisenstein2001}
{Eisenstein} D.~J. {et~al.}, 2001, \aj, 122, 2267

\bibitem[{Elvin-Poole {et~al}\mbox{.}(2018)Elvin-Poole, Crocce, Ross,
  Giannantonio, Rozo, Rykoff, Avila, Banik, Blazek, Bridle, Cawthon,
  Drlica-Wagner, Friedrich, Kokron, Krause, MacCrann, Prat, S\'anchez, Secco,
  Sevilla-Noarbe, Troxel, Abbott, Abdalla, Allam, Annis, Asorey, Bechtol,
  Becker, Benoit-L\'evy, Bernstein, Bertin, Brooks, Buckley-Geer, Burke,
  Carnero~Rosell, Carollo, Carrasco~Kind, Carretero, Castander, Cunha,
  D'Andrea, da~Costa, Davis, Davis, Desai, Diehl, Dietrich, Dodelson, Doel,
  Eifler, Evrard, Fernandez, Flaugher, Fosalba, Frieman, Garc\'{\i}a-Bellido,
  Gaztanaga, Gerdes, Glazebrook, Gruen, Gruendl, Gschwend, Gutierrez, Hartley,
  Hinton, Honscheid, Hoormann, Jain, James, Jarvis, Jeltema, Johnson, Johnson,
  King, Kuehn, Kuhlmann, Kuropatkin, Lahav, Lewis, Li, Lidman, Lima, Lin,
  Macaulay, March, Marshall, Martini, Melchior, Menanteau, Miquel, Mohr,
  M\"oller, Nichol, Nord, O'Neill, Percival, Petravick, Plazas, Romer, Sako,
  Sanchez, Scarpine, Schindler, Schubnell, Sheldon, Smith, Smith,
  Soares-Santos, Sobreira, Sommer, Suchyta, Swanson, Tarle, Thomas, Tucker,
  Tucker, Uddin, Vikram, Walker, Wechsler, Weller, Wester, Wolf, Yuan, Zhang,
  \& Zuntz}]{ElvinPoole2017}
Elvin-Poole J. {et~al.}, 2018, Phys. Rev. D, 98, 042006

\bibitem[{{Eriksen} \& {Gazta{\~n}aga}(2015)}]{Eriksen2015}
{Eriksen} M., {Gazta{\~n}aga} E., 2015, \mnras, 452, 2168

\bibitem[{{Fitzpatrick} \& {Massa}(1986)}]{Fitzpatrick1986}
{Fitzpatrick} E.~L., {Massa} D., 1986, \apj, 307, 286

\bibitem[{{Fitzpatrick} \& {Massa}(2007)}]{Fitzpatrick2007}
{Fitzpatrick} E.~L., {Massa} D., 2007, \apj, 663, 320

\bibitem[{{Gaia Collaboration}(2016)}]{Brown2016}
{Gaia Collaboration}, 2016, \aap, 595, A2

\bibitem[{{Gazta{\~n}aga}, {Cabr{\'e}} \& {Hui}(2009){Gazta{\~n}aga},
  {Cabr{\'e}}, \& {Hui}}]{Gaztanaga2009}
{Gazta{\~n}aga} E., {Cabr{\'e}} A., {Hui} L., 2009, \mnras, 399, 1663

\bibitem[{{Gazta{\~n}aga} {et~al}\mbox{.}(2012){Gazta{\~n}aga}, {Eriksen},
  {Crocce}, {Castander}, {Fosalba}, {Marti}, {Miquel}, \&
  {Cabr{\'e}}}]{Gaztanaga2012}
{Gazta{\~n}aga} E., {Eriksen} M., {Crocce} M., {Castander} F.~J., {Fosalba} P.,
  {Marti} P., {Miquel} R., {Cabr{\'e}} A., 2012, \mnras, 422, 2904

\bibitem[{{Heavens}, {Refregier} \& {Heymans}(2000){Heavens}, {Refregier}, \&
  {Heymans}}]{Heavens2000}
{Heavens} A., {Refregier} A., {Heymans} C., 2000, \mnras, 319, 649

\bibitem[{{Heymans} {et~al}\mbox{.}(2012){Heymans}, {Van Waerbeke}, {Miller},
  {Erben}, {Hildebrandt}, {Hoekstra}, {Kitching}, {Mellier}, {Simon},
  {Bonnett}, {Coupon}, {Fu}, {Harnois D{\'e}raps}, {Hudson}, {Kilbinger},
  {Kuijken}, {Rowe}, {Schrabback}, {Semboloni}, {van Uitert}, {Vafaei}, \&
  {Velander}}]{Heymans2012}
{Heymans} C. {et~al.}, 2012, \mnras, 427, 146

\bibitem[{{Hildebrandt} {et~al}\mbox{.}(2012){Hildebrandt}, {Erben}, {Kuijken},
  {van Waerbeke}, {Heymans}, {Coupon}, {Benjamin}, {Bonnett}, {Fu}, {Hoekstra},
  {Kitching}, {Mellier}, {Miller}, {Velander}, {Hudson}, {Rowe}, {Schrabback},
  {Semboloni}, \& {Ben{\'{\i}}tez}}]{Hildebrandt2012}
{Hildebrandt} H. {et~al.}, 2012, \mnras, 421, 2355

\bibitem[{{Hirata} \& {Seljak}(2004)}]{Hirata2004}
{Hirata} C.~M., {Seljak} U., 2004, \prd, 70, 063526

\bibitem[{{Hogg} {et~al}\mbox{.}(2002){Hogg}, {Baldry}, {Blanton}, \&
  {Eisenstein}}]{Hogg2002}
{Hogg} D.~W., {Baldry} I.~K., {Blanton} M.~R., {Eisenstein} D.~J., 2002,
  preprint(arXiv:astro-ph/0210394)

\bibitem[{Hoyer \& Hamman(2017)}]{Hoyer2017}
Hoyer S., Hamman J., 2017, Journal of Open Research Software, 5

\bibitem[{{Hoyle} {et~al}\mbox{.}(2018){Hoyle}, {Gruen}, {Bernstein}, {Rau},
  {De Vicente}, {Hartley}, {Gaztanaga}, {DeRose}, {Troxel}, {Davis}, {Alarcon},
  {MacCrann}, {Prat}, {S{\'a}nchez}, {Sheldon}, {Wechsler}, {Asorey}, {Becker},
  {Bonnett}, {Carnero Rosell}, {Carollo}, {Carrasco Kind}, {Castander},
  {Cawthon}, {Chang}, {Childress}, {Davis}, {Drlica-Wagner}, {Gatti},
  {Glazebrook}, {Gschwend}, {Hinton}, {Hoormann}, {Kim}, {King}, {Kuehn},
  {Lewis}, {Lidman}, {Lin}, {Macaulay}, {Maia}, {Martini}, {Mudd},
  {M{\"o}ller}, {Nichol}, {Ogando}, {Rollins}, {Roodman}, {Ross}, {Rozo},
  {Rykoff}, {Samuroff}, {Sevilla-Noarbe}, {Sharp}, {Sommer}, {Tucker}, {Uddin},
  {Varga}, {Vielzeuf}, {Yuan}, {Zhang}, {Abbott}, {Abdalla}, {Allam}, {Annis},
  {Bechtol}, {Benoit-L{\'e}vy}, {Bertin}, {Brooks}, {Buckley-Geer}, {Burke},
  {Busha}, {Capozzi}, {Carretero}, {Crocce}, {D'Andrea}, {da Costa}, {DePoy},
  {Desai}, {Diehl}, {Doel}, {Eifler}, {Estrada}, {Evrard}, {Fernandez},
  {Flaugher}, {Fosalba}, {Frieman}, {Garc{\'{\i}}a-Bellido}, {Gerdes},
  {Giannantonio}, {Goldstein}, {Gruendl}, {Gutierrez}, {Honscheid}, {James},
  {Jarvis}, {Jeltema}, {Johnson}, {Johnson}, {Kirk}, {Krause}, {Kuhlmann},
  {Kuropatkin}, {Lahav}, {Li}, {Lima}, {March}, {Marshall}, {Melchior},
  {Menanteau}, {Miquel}, {Nord}, {O'Neill}, {Plazas}, {Romer}, {Sako},
  {Sanchez}, {Santiago}, {Scarpine}, {Schindler}, {Schubnell}, {Smith},
  {Smith}, {Soares-Santos}, {Sobreira}, {Suchyta}, {Swanson}, {Tarle},
  {Thomas}, {Tucker}, {Vikram}, {Walker}, {Weller}, {Wester}, {Wolf}, {Yanny},
  \& {Zuntz}}]{Hoyle2018}
{Hoyle} B. {et~al.}, 2018, \mnras, 478, 592

\bibitem[{{Ilbert} {et~al}\mbox{.}(2009){Ilbert}, {Capak}, {Salvato}, {Aussel},
  {McCracken}, {Sanders}, {Scoville}, {Kartaltepe}, {Arnouts}, {Le Floc'h},
  {Mobasher}, {Taniguchi}, {Lamareille}, {Leauthaud}, {Sasaki}, {Thompson},
  {Zamojski}, {Zamorani}, {Bardelli}, {Bolzonella}, {Bongiorno}, {Brusa},
  {Caputi}, {Carollo}, {Contini}, {Cook}, {Coppa}, {Cucciati}, {de la Torre},
  {de Ravel}, {Franzetti}, {Garilli}, {Hasinger}, {Iovino}, {Kampczyk},
  {Kneib}, {Knobel}, {Kovac}, {Le Borgne}, {Le Brun}, {Le F{\`e}vre}, {Lilly},
  {Looper}, {Maier}, {Mainieri}, {Mellier}, {Mignoli}, {Murayama}, {Pell{\`o}},
  {Peng}, {P{\'e}rez-Montero}, {Renzini}, {Ricciardelli}, {Schiminovich},
  {Scodeggio}, {Shioya}, {Silverman}, {Surace}, {Tanaka}, {Tasca}, {Tresse},
  {Vergani}, \& {Zucca}}]{Ilbert2009}
{Ilbert} O. {et~al.}, 2009, \apj, 690, 1236

\bibitem[{{Ilbert} {et~al}\mbox{.}(2013){Ilbert}, {McCracken}, {Le F{\`e}vre},
  {Capak}, {Dunlop}, {Karim}, {Renzini}, {Caputi}, {Boissier}, {Arnouts},
  {Aussel}, {Comparat}, {Guo}, {Hudelot}, {Kartaltepe}, {Kneib}, {Krogager},
  {Le Floc'h}, {Lilly}, {Mellier}, {Milvang-Jensen}, {Moutard}, {Onodera},
  {Richard}, {Salvato}, {Sanders}, {Scoville}, {Silverman}, {Taniguchi},
  {Tasca}, {Thomas}, {Toft}, {Tresse}, {Vergani}, {Wolk}, \&
  {Zirm}}]{Ilbert2013}
{Ilbert} O. {et~al.}, 2013, \aap, 556, A55

\bibitem[{{Joachimi} {et~al}\mbox{.}(2015){Joachimi}, {Cacciato}, {Kitching},
  {Leonard}, {Mandelbaum}, {Sch{\"a}fer}, {Sif{\'o}n}, {Hoekstra}, {Kiessling},
  {Kirk}, \& {Rassat}}]{Joachimi2015}
{Joachimi} B. {et~al.}, 2015, \ssr, 193, 1

\bibitem[{{Joachimi} {et~al}\mbox{.}(2011){Joachimi}, {Mandelbaum}, {Abdalla},
  \& {Bridle}}]{Joachimi2011}
{Joachimi} B., {Mandelbaum} R., {Abdalla} F.~B., {Bridle} S.~L., 2011, \aap,
  527, A26

\bibitem[{{Jouvel} {et~al}\mbox{.}(2014){Jouvel}, {Abdalla}, {Kirk}, {Lahav},
  {Lin}, {Annis}, {Kron}, \& {Frieman}}]{Jouvel2014}
{Jouvel} S., {Abdalla} F.~B., {Kirk} D., {Lahav} O., {Lin} H., {Annis} J.,
  {Kron} R., {Frieman} J.~A., 2014, \mnras, 438, 2218

\bibitem[{{Knobel} {et~al}\mbox{.}(2012){Knobel}, {Lilly}, {Iovino}, {Kova{\v
  c}}, {Bschorr}, {Presotto}, {Oesch}, {Kampczyk}, {Carollo}, {Contini},
  {Kneib}, {Le Fevre}, {Mainieri}, {Renzini}, {Scodeggio}, {Zamorani},
  {Bardelli}, {Bolzonella}, {Bongiorno}, {Caputi}, {Cucciati}, {de la Torre},
  {de Ravel}, {Franzetti}, {Garilli}, {Lamareille}, {Le Borgne}, {Le Brun},
  {Maier}, {Mignoli}, {Pello}, {Peng}, {Perez Montero}, {Silverman}, {Tanaka},
  {Tasca}, {Tresse}, {Vergani}, {Zucca}, {Barnes}, {Bordoloi}, {Cappi},
  {Cimatti}, {Coppa}, {Koekemoer}, {L{\'o}pez-Sanjuan}, {McCracken}, {Moresco},
  {Nair}, {Pozzetti}, \& {Welikala}}]{Knobel2012}
{Knobel} C. {et~al.}, 2012, \apj, 753, 121

\bibitem[{{Koekemoer} {et~al}\mbox{.}(2007){Koekemoer}, {Aussel}, {Calzetti},
  {Capak}, {Giavalisco}, {Kneib}, {Leauthaud}, {Le F{\`e}vre}, {McCracken},
  {Massey}, {Mobasher}, {Rhodes}, {Scoville}, \& {Shopbell}}]{Koekemoer2007}
{Koekemoer} A.~M. {et~al.}, 2007, \apjs, 172, 196

\bibitem[{{Kuijken} {et~al}\mbox{.}(2015){Kuijken}, {Heymans}, {Hildebrandt},
  {Nakajima}, {Erben}, {de Jong}, {Viola}, {Choi}, {Hoekstra}, {Miller}, {van
  Uitert}, {Amon}, {Blake}, {Brouwer}, {Buddendiek}, {Conti}, {Eriksen},
  {Grado}, {Harnois-D{\'e}raps}, {Helmich}, {Herbonnet}, {Irisarri},
  {Kitching}, {Klaes}, {La Barbera}, {Napolitano}, {Radovich}, {Schneider},
  {Sif{\'o}n}, {Sikkema}, {Simon}, {Tudorica}, {Valentijn}, {Verdoes Kleijn},
  \& {van Waerbeke}}]{Kuijken2015}
{Kuijken} K. {et~al.}, 2015, \mnras, 454, 3500

\bibitem[{{Laigle} {et~al}\mbox{.}(2016){Laigle}, {McCracken}, {Ilbert},
  {Hsieh}, {Davidzon}, {Capak}, {Hasinger}, {Silverman}, {Pichon}, {Coupon},
  {Aussel}, {Le Borgne}, {Caputi}, {Cassata}, {Chang}, {Civano}, {Dunlop},
  {Fynbo}, {Kartaltepe}, {Koekemoer}, {Le F{\`e}vre}, {Le Floc'h}, {Leauthaud},
  {Lilly}, {Lin}, {Marchesi}, {Milvang-Jensen}, {Salvato}, {Sanders},
  {Scoville}, {Smolcic}, {Stockmann}, {Taniguchi}, {Tasca}, {Toft}, {Vaccari},
  \& {Zabl}}]{Laigle2016}
{Laigle} C. {et~al.}, 2016, \apjs, 224, 24

\bibitem[{Lam, Pitrou \& Seibert(2015)Lam, Pitrou, \& Seibert}]{Lam2015}
Lam S.~K., Pitrou A., Seibert S., 2015, in Proceedings of the Second Workshop
  on the LLVM Compiler Infrastructure in HPC, LLVM '15, ACM, New York, NY, USA,
  pp. 7:1--7:6

\bibitem[{{Laureijs} {et~al}\mbox{.}(2011){Laureijs}, {Amiaux}, {Arduini},
  {Augu{\`e}res}, {Brinchmann}, {Cole}, {Cropper}, {Dabin}, {Duvet}, {Ealet},
  \& et~al.}]{euclidrb}
{Laureijs} R. {et~al.}, 2011, preprint(arXiv:1110.3193)

\bibitem[{{Leauthaud} {et~al}\mbox{.}(2007){Leauthaud}, {Massey}, {Kneib},
  {Rhodes}, {Johnston}, {Capak}, {Heymans}, {Ellis}, {Koekemoer}, {Le
  F{\`e}vre}, {Mellier}, {R{\'e}fr{\'e}gier}, {Robin}, {Scoville}, {Tasca},
  {Taylor}, \& {Van Waerbeke}}]{Leauthaud2007}
{Leauthaud} A. {et~al.}, 2007, \apjs, 172, 219

\bibitem[{{Lilly} {et~al}\mbox{.}(2009){Lilly}, {Le Brun}, {Maier}, {Mainieri},
  {Mignoli}, {Scodeggio}, {Zamorani}, {Carollo}, {Contini}, {Kneib}, {Le
  F{\`e}vre}, {Renzini}, {Bardelli}, {Bolzonella}, {Bongiorno}, {Caputi},
  {Coppa}, {Cucciati}, {de la Torre}, {de Ravel}, {Franzetti}, {Garilli},
  {Iovino}, {Kampczyk}, {Kovac}, {Knobel}, {Lamareille}, {Le Borgne}, {Pello},
  {Peng}, {P{\'e}rez-Montero}, {Ricciardelli}, {Silverman}, {Tanaka}, {Tasca},
  {Tresse}, {Vergani}, {Zucca}, {Ilbert}, {Salvato}, {Oesch}, {Abbas},
  {Bottini}, {Capak}, {Cappi}, {Cassata}, {Cimatti}, {Elvis}, {Fumana},
  {Guzzo}, {Hasinger}, {Koekemoer}, {Leauthaud}, {Maccagni}, {Marinoni},
  {McCracken}, {Memeo}, {Meneux}, {Porciani}, {Pozzetti}, {Sanders},
  {Scaramella}, {Scarlata}, {Scoville}, {Shopbell}, \& {Taniguchi}}]{Lilly2009}
{Lilly} S.~J. {et~al.}, 2009, \apjs, 184, 218

\bibitem[{{Lilly} {et~al}\mbox{.}(2007){Lilly}, {Le F{\`e}vre}, {Renzini},
  {Zamorani}, {Scodeggio}, {Contini}, {Carollo}, {Hasinger}, {Kneib}, {Iovino},
  {Le Brun}, {Maier}, {Mainieri}, {Mignoli}, {Silverman}, {Tasca},
  {Bolzonella}, {Bongiorno}, {Bottini}, {Capak}, {Caputi}, {Cimatti},
  {Cucciati}, {Daddi}, {Feldmann}, {Franzetti}, {Garilli}, {Guzzo}, {Ilbert},
  {Kampczyk}, {Kovac}, {Lamareille}, {Leauthaud}, {Le Borgne}, {McCracken},
  {Marinoni}, {Pello}, {Ricciardelli}, {Scarlata}, {Vergani}, {Sanders},
  {Schinnerer}, {Scoville}, {Taniguchi}, {Arnouts}, {Aussel}, {Bardelli},
  {Brusa}, {Cappi}, {Ciliegi}, {Finoguenov}, {Foucaud}, {Franceschini},
  {Halliday}, {Impey}, {Knobel}, {Koekemoer}, {Kurk}, {Maccagni}, {Maddox},
  {Marano}, {Marconi}, {Meneux}, {Mobasher}, {Moreau}, {Peacock}, {Porciani},
  {Pozzetti}, {Scaramella}, {Schiminovich}, {Shopbell}, {Smail}, {Thompson},
  {Tresse}, {Vettolani}, {Zanichelli}, \& {Zucca}}]{Lilly2007}
{Lilly} S.~J. {et~al.}, 2007, \apjs, 172, 70

\bibitem[{{LSST Science Collaboration}(2009)}]{lsst1}
{LSST Science Collaboration}, 2009, preprint(arXiv:0912.0201)

\bibitem[{{Madrid} {et~al}\mbox{.}(2010){Madrid}, {Ballester}, {Cardiel-Sas},
  {Casas}, {Castander}, {Castilla}, {de Vicente}, {Fernández}, {Gaztañaga},
  {Grañena}, {Jiménez}, {Maiorino}, {Martí}, {Miquel}, {Sánchez},
  {Serrano}, {Sevilla}, \& {Tonello}}]{Madrid2010}
{Madrid} F. {et~al.}, 2010, in Proc.SPIE, Vol. 7735, pp. 7735 -- 7735 -- 7

\bibitem[{{Mandelbaum} {et~al}\mbox{.}(2013){Mandelbaum}, {Slosar}, {Baldauf},
  {Seljak}, {Hirata}, {Nakajima}, {Reyes}, \& {Smith}}]{Mandelbaum2013}
{Mandelbaum} R., {Slosar} A., {Baldauf} T., {Seljak} U., {Hirata} C.~M.,
  {Nakajima} R., {Reyes} R., {Smith} R.~E., 2013, \mnras, 432, 1544

\bibitem[{{Mart{\'{\i}}} {et~al}\mbox{.}(2014{\natexlab{a}}){Mart{\'{\i}}},
  {Miquel}, {Bauer}, \& {Gazta{\~n}aga}}]{Marti2014b}
{Mart{\'{\i}}} P., {Miquel} R., {Bauer} A., {Gazta{\~n}aga} E.,
  2014{\natexlab{a}}, \mnras, 437, 3490

\bibitem[{{Mart{\'{\i}}} {et~al}\mbox{.}(2014{\natexlab{b}}){Mart{\'{\i}}},
  {Miquel}, {Castander}, {Gazta{\~n}aga}, {Eriksen}, \&
  {S{\'a}nchez}}]{Marti2014}
{Mart{\'{\i}}} P., {Miquel} R., {Castander} F.~J., {Gazta{\~n}aga} E.,
  {Eriksen} M., {S{\'a}nchez} C., 2014{\natexlab{b}}, \mnras, 442, 92

\bibitem[{McKinney(2010)}]{McKinney2010}
McKinney W., 2010, in Proceedings of the 9th Python in Science Conference,
  van~der Walt S., Millman J., eds., pp. 51 -- 56

\bibitem[{{Moles} {et~al}\mbox{.}(2008){Moles}, {Ben{\'{\i}}tez}, {Aguerri},
  {Alfaro}, {Broadhurst}, {Cabrera-Ca{\~n}o}, {Castander}, {Cepa},
  {Cervi{\~n}o}, {Crist{\'o}bal-Hornillos}, {Fern{\'a}ndez-Soto}, {Gonz{\'a}lez
  Delgado}, {Infante}, {M{\'a}rquez}, {Mart{\'{\i}}nez}, {Masegosa}, {del
  Olmo}, {Perea}, {Prada}, {Quintana}, \& {S{\'a}nchez}}]{Moles2008}
{Moles} M. {et~al.}, 2008, \aj, 136, 1325

\bibitem[{{Molino} {et~al}\mbox{.}(2014){Molino}, {Ben{\'{\i}}tez}, {Moles},
  {Fern{\'a}ndez-Soto}, {Crist{\'o}bal-Hornillos}, {Ascaso},
  {Jim{\'e}nez-Teja}, {Schoenell}, {Arnalte-Mur}, {Povi{\'c}}, {Coe},
  {L{\'o}pez-Sanjuan}, {D{\'{\i}}az-Garc{\'{\i}}a}, {Varela}, {Stefanon},
  {Cenarro}, {Matute}, {Masegosa}, {M{\'a}rquez}, {Perea}, {Del Olmo},
  {Husillos}, {Alfaro}, {Aparicio-Villegas}, {Cervi{\~n}o}, {Huertas-Company},
  {Aguerri}, {Broadhurst}, {Cabrera-Ca{\~n}o}, {Cepa}, {Gonz{\'a}lez},
  {Infante}, {Mart{\'{\i}}nez}, {Prada}, \& {Quintana}}]{Molino2014}
{Molino} A. {et~al.}, 2014, \mnras, 441, 2891

\bibitem[{{Nakajima} {et~al}\mbox{.}(2012){Nakajima}, {Mandelbaum}, {Seljak},
  {Cohn}, {Reyes}, \& {Cool}}]{Nakajima2012}
{Nakajima} R., {Mandelbaum} R., {Seljak} U., {Cohn} J.~D., {Reyes} R., {Cool}
  R., 2012, \mnras, 420, 3240

\bibitem[{{Padilla} {et~al}\mbox{.}(in prep.){Padilla}, {Castander},
  {Fernandez}, \& {et al.}}]{PAUcam}
{Padilla} C., {Castander} F., {Fernandez} E., {et al.}, in prep.

\bibitem[{{Pickles}(1998)}]{Pickles1998}
{Pickles} A.~J., 1998, \pasp, 110, 863

\bibitem[{{Planck Collaboration} {et~al}\mbox{.}(2016){Planck Collaboration},
  {Ade}, {Aghanim}, {Arnaud}, {Ashdown}, {Aumont}, {Baccigalupi}, {Banday},
  {Barreiro}, {Bartlett}, \& et~al.}]{Planck2015}
{Planck Collaboration} {et~al.}, 2016, \aap, 594, A13

\bibitem[{Prat {et~al}\mbox{.}(2018)Prat, S\'anchez, Fang, Gruen, Elvin-Poole,
  Kokron, Secco, Jain, Miquel, MacCrann, Troxel, Alarcon, Bacon, Bernstein,
  Blazek, Cawthon, Chang, Crocce, Davis, De~Vicente, Dietrich, Drlica-Wagner,
  Friedrich, Gatti, Hartley, Hoyle, Huff, Jarvis, Rau, Rollins, Ross, Rozo,
  Rykoff, Samuroff, Sheldon, Varga, Vielzeuf, Zuntz, Abbott, Abdalla, Allam,
  Annis, Bechtol, Benoit-L\'evy, Bertin, Brooks, Buckley-Geer, Burke,
  Carnero~Rosell, Carrasco~Kind, Carretero, Castander, Cunha, D'Andrea,
  da~Costa, Desai, Diehl, Dodelson, Eifler, Fernandez, Flaugher, Fosalba,
  Frieman, Garc\'{\i}a-Bellido, Gaztanaga, Gerdes, Giannantonio, Goldstein,
  Gruendl, Gschwend, Gutierrez, Honscheid, James, Jeltema, Johnson, Johnson,
  Kirk, Krause, Kuehn, Kuhlmann, Lahav, Li, Lima, Maia, March, Marshall,
  Martini, Melchior, Menanteau, Mohr, Nichol, Nord, Plazas, Romer, Roodman,
  Sako, Sanchez, Scarpine, Schindler, Schubnell, Sevilla-Noarbe, Smith, Smith,
  Soares-Santos, Sobreira, Suchyta, Swanson, Tarle, Thomas, Tucker, Vikram,
  Walker, Wechsler, Yanny, \& Zhang}]{Prat2017}
Prat J. {et~al.}, 2018, Phys. Rev. D, 98, 042005

\bibitem[{{Prevot} {et~al}\mbox{.}(1984){Prevot}, {Lequeux}, {Maurice},
  {Prevot}, \& {Rocca-Volmerange}}]{Prevot1984}
{Prevot} M.~L., {Lequeux} J., {Maurice} E., {Prevot} L., {Rocca-Volmerange} B.,
  1984, \aap, 132, 389

\bibitem[{{Rozo} {et~al}\mbox{.}(2016){Rozo}, {Rykoff}, {Abate}, {Bonnett},
  {Crocce}, {Davis}, {Hoyle}, {Leistedt}, {Peiris}, {Wechsler}, {Abbott},
  {Abdalla}, {Banerji}, {Bauer}, {Benoit-L{\'e}vy}, {Bernstein}, {Bertin},
  {Brooks}, {Buckley-Geer}, {Burke}, {Capozzi}, {Rosell}, {Carollo}, {Kind},
  {Carretero}, {Castander}, {Childress}, {Cunha}, {D'Andrea}, {Davis}, {DePoy},
  {Desai}, {Diehl}, {Dietrich}, {Doel}, {Eifler}, {Evrard}, {Neto}, {Flaugher},
  {Fosalba}, {Frieman}, {Gaztanaga}, {Gerdes}, {Glazebrook}, {Gruen},
  {Gruendl}, {Honscheid}, {James}, {Jarvis}, {Kim}, {Kuehn}, {Kuropatkin},
  {Lahav}, {Lidman}, {Lima}, {Maia}, {March}, {Martini}, {Melchior}, {Miller},
  {Miquel}, {Mohr}, {Nichol}, {Nord}, {O'Neill}, {Ogando}, {Plazas}, {Romer},
  {Roodman}, {Sako}, {Sanchez}, {Santiago}, {Schubnell}, {Sevilla-Noarbe},
  {Smith}, {Soares-Santos}, {Sobreira}, {Suchyta}, {Swanson}, {Thaler},
  {Thomas}, {Uddin}, {Vikram}, {Walker}, {Wester}, {Zhang}, \& {da
  Costa}}]{Rozo2016}
{Rozo} E. {et~al.}, 2016, \mnras, 461, 1431

\bibitem[{{Sadeh}, {Abdalla} \& {Lahav}(2016){Sadeh}, {Abdalla}, \&
  {Lahav}}]{Sadeh2016}
{Sadeh} I., {Abdalla} F.~B., {Lahav} O., 2016, \pasp, 128, 104502

\bibitem[{{Sargent} {et~al}\mbox{.}(2007){Sargent}, {Carollo}, {Lilly},
  {Scarlata}, {Feldmann}, {Kampczyk}, {Koekemoer}, {Scoville}, {Kneib},
  {Leauthaud}, {Massey}, {Rhodes}, {Tasca}, {Capak}, {McCracken}, {Porciani},
  {Renzini}, {Taniguchi}, {Thompson}, \& {Sheth}}]{Sargent2007}
{Sargent} M.~T. {et~al.}, 2007, \apjs, 172, 434

\bibitem[{{Schlegel}, {Finkbeiner} \& {Davis}(1998){Schlegel}, {Finkbeiner}, \&
  {Davis}}]{Schlegel1998}
{Schlegel} D.~J., {Finkbeiner} D.~P., {Davis} M., 1998, \apj, 500, 525

\bibitem[{{Serrano} {et~al}\mbox{.}(in prep.){Serrano}, {Castander},
  {Fernandez}, \& {et al.}}]{PAUimage}
{Serrano} S., {Castander} F., {Fernandez} E., {et al.}, in prep.

\bibitem[{{Serrano} {et~al}\mbox{.}({in prep.}){Serrano}, {Gazta{\~n}aga},
  {Eriksen}, \& {Castander}}]{PAUphoto}
{Serrano} S., {Gazta{\~n}aga} E., {Eriksen} M., {Castander} F., {in prep.}

\bibitem[{Sha {et~al}\mbox{.}(2007)Sha, Lin, Saul, \& Lee}]{Sha2007}
Sha F., Lin Y., Saul L.~K., Lee D.~D., 2007, Neural Comput., 19, 2004

\bibitem[{{Smith} {et~al}\mbox{.}(2002){Smith}, {Tucker}, {Kent}, {Richmond},
  {Fukugita}, {Ichikawa}, {Ichikawa}, {Jorgensen}, {Uomoto}, {Gunn}, {Hamabe},
  {Watanabe}, {Tolea}, {Henden}, {Annis}, {Pier}, {McKay}, {Brinkmann}, {Chen},
  {Holtzman}, {Shimasaku}, \& {York}}]{Smith2002}
{Smith} J.~A. {et~al.}, 2002, \aj, 123, 2121

\bibitem[{{Spergel} {et~al}\mbox{.}(2015){Spergel}, {Gehrels}, {Baltay},
  {Bennett}, {Breckinridge}, {Donahue}, {Dressler}, {Gaudi}, {Greene}, {Guyon},
  {Hirata}, {Kalirai}, {Kasdin}, {Macintosh}, {Moos}, {Perlmutter}, {Postman},
  {Rauscher}, {Rhodes}, {Wang}, {Weinberg}, {Benford}, {Hudson}, {Jeong},
  {Mellier}, {Traub}, {Yamada}, {Capak}, {Colbert}, {Masters}, {Penny},
  {Savransky}, {Stern}, {Zimmerman}, {Barry}, {Bartusek}, {Carpenter}, {Cheng},
  {Content}, {Dekens}, {Demers}, {Grady}, {Jackson}, {Kuan}, {Kruk}, {Melton},
  {Nemati}, {Parvin}, {Poberezhskiy}, {Peddie}, {Ruffa}, {Wallace}, {Whipple},
  {Wollack}, \& {Zhao}}]{wfirst}
{Spergel} D. {et~al.}, 2015, preprint(arXiv:1503.03757)

\bibitem[{{Stecher} \& {Donn}(1965)}]{Stecher1965}
{Stecher} T.~P., {Donn} B., 1965, \apj, 142, 1681

\bibitem[{{Strauss} {et~al}\mbox{.}(2002){Strauss}, {Weinberg}, {Lupton},
  {Narayanan}, {Annis}, {Bernardi}, {Blanton}, {Burles}, {Connolly},
  {Dalcanton}, {Doi}, {Eisenstein}, {Frieman}, {Fukugita}, {Gunn},
  {Ivezi{\'c}}, {Kent}, {Kim}, {Knapp}, {Kron}, {Munn}, {Newberg}, {Nichol},
  {Okamura}, {Quinn}, {Richmond}, {Schlegel}, {Shimasaku}, {SubbaRao},
  {Szalay}, {Vanden Berk}, {Vogeley}, {Yanny}, {Yasuda}, {York}, \&
  {Zehavi}}]{Strauss2002}
{Strauss} M.~A. {et~al.}, 2002, \aj, 124, 1810

\bibitem[{Stroustrup(2000)}]{Stroustrup2000}
Stroustrup B., 2000, The C++ Programming Language, 3rd edn. Addison-Wesley
  Longman Publishing Co., Inc., Boston, MA, USA

\bibitem[{{Tanaka} {et~al}\mbox{.}(2018){Tanaka}, {Coupon}, {Hsieh}, {Mineo},
  {Nishizawa}, {Speagle}, {Furusawa}, {Miyazaki}, \& {Murayama}}]{Tanaka2018}
{Tanaka} M. {et~al.}, 2018, \pasj, 70, S9

\bibitem[{Tanaka {et~al}\mbox{.}(2004)Tanaka, Goto, Okamura, Shimasaku, \&
  Brinkmann}]{Tanaka2004}
Tanaka M., Goto T., Okamura S., Shimasaku K., Brinkmann J., 2004, The
  Astronomical Journal, 128, 2677

\bibitem[{{Taniguchi} {et~al}\mbox{.}(2015){Taniguchi}, {Kajisawa},
  {Kobayashi}, {Shioya}, {Nagao}, {Capak}, {Aussel}, {Ichikawa}, {Murayama},
  {Scoville}, {Ilbert}, {Salvato}, {Sanders}, {Mobasher}, {Miyazaki},
  {Komiyama}, {Le F{\`e}vre}, {Tasca}, {Lilly}, {Carollo}, {Renzini}, {Rich},
  {Schinnerer}, {Kaifu}, {Karoji}, {Arimoto}, {Okamura}, {Ohta}, {Shimasaku},
  \& {Hayashino}}]{Taniguchi2015}
{Taniguchi} Y. {et~al.}, 2015, \pasj, 67, 104

\bibitem[{{Tegmark} {et~al}\mbox{.}(2006){Tegmark}, {Eisenstein}, {Strauss},
  {Weinberg}, {Blanton}, {Frieman}, {Fukugita}, {Gunn}, {Hamilton}, {Knapp},
  {Nichol}, {Ostriker}, {Padmanabhan}, {Percival}, {Schlegel}, {Schneider},
  {Scoccimarro}, {Seljak}, {Seo}, {Swanson}, {Szalay}, {Vogeley}, {Yoo},
  {Zehavi}, {Abazajian}, {Anderson}, {Annis}, {Bahcall}, {Bassett}, {Berlind},
  {Brinkmann}, {Budavari}, {Castander}, {Connolly}, {Csabai}, {Doi},
  {Finkbeiner}, {Gillespie}, {Glazebrook}, {Hennessy}, {Hogg}, {Ivezi{\'c}},
  {Jain}, {Johnston}, {Kent}, {Lamb}, {Lee}, {Lin}, {Loveday}, {Lupton},
  {Munn}, {Pan}, {Park}, {Peoples}, {Pier}, {Pope}, {Richmond}, {Rockosi},
  {Scranton}, {Sheth}, {Stebbins}, {Stoughton}, {Szapudi}, {Tucker}, {vanden
  Berk}, {Yanny}, \& {York}}]{Tegmark2006}
{Tegmark} M. {et~al.}, 2006, \prd, 74, 123507

\bibitem[{{The Dark Energy Survey Collaboration}(2005)}]{DES2005}
{The Dark Energy Survey Collaboration}, 2005, preprint (ArXiv:astro-ph/0510346)

\bibitem[{{Tonello} {et~al}\mbox{.}(in prep.){Tonello}, {Tallada}, {Serrano},
  \& {et al.}}]{PAUdm}
{Tonello} N., {Tallada} P., {Serrano} S., {et al.}, in prep.

\bibitem[{{Tortorelli} {et~al}\mbox{.}(2018){Tortorelli}, {Della Bruna},
  {Herbel}, {Amara}, {Refregier}, {Alarcon}, {Castander}, {De Vicente},
  {Eriksen}, {Fernandez}, {Garc{\'{\i}}a-Bellido}, {Gaztanaga}, {Miquel},
  {Padilla}, {Sanchez}, {Serrano}, {Stothert}, \& {Tonello}}]{Tortorelli2018}
{Tortorelli} L. {et~al.}, 2018, preprint(ArXiv:1805.05340)

\bibitem[{{Troxel} \& {Ishak}(2015)}]{Troxel2015}
{Troxel} M.~A., {Ishak} M., 2015, \physrep, 558, 1

\bibitem[{{van Dokkum}(2001)}]{Dokkum2001}
{van Dokkum} P.~G., 2001, \pasp, 113, 1420

\bibitem[{{van Rossum}(1995)}]{Rossum1995}
{van Rossum} G., 1995, {Python tutorial, Technical Report CS-R9526}. Tech.
  rep., Centrum voor Wiskunde en Informatica (CWI), Amsterdam

\bibitem[{{van Uitert} {et~al}\mbox{.}(2015){van Uitert}, {Cacciato},
  {Hoekstra}, \& {Herbonnet}}]{vanUitert2015}
{van Uitert} E., {Cacciato} M., {Hoekstra} H., {Herbonnet} R., 2015, \aap, 579,
  A26

\bibitem[{{Weinberg} {et~al}\mbox{.}(2013){Weinberg}, {Mortonson},
  {Eisenstein}, {Hirata}, {Riess}, \& {Rozo}}]{Weinberg2013}
{Weinberg} D.~H., {Mortonson} M.~J., {Eisenstein} D.~J., {Hirata} C., {Riess}
  A.~G., {Rozo} E., 2013, \physrep, 530, 87

\bibitem[{{Xiang}, {Li} \& {Zhong}(2011){Xiang}, {Li}, \& {Zhong}}]{Xiang2011}
{Xiang} F.~Y., {Li} A., {Zhong} J.~X., 2011, \apj, 733, 91

\bibitem[{Zaharia {et~al}\mbox{.}(2010)Zaharia, Chowdhury, Franklin, Shenker,
  \& Stoica}]{Zaharia2010}
Zaharia M., Chowdhury M., Franklin M.~J., Shenker S., Stoica I., 2010, in
  Proceedings of the 2Nd USENIX Conference on Hot Topics in Cloud Computing,
  HotCloud'10, USENIX Association, Berkeley, CA, USA, pp. 10--10

\end{thebibliography}
\bibliographystyle{mn2e}
\end{document}